\documentclass[sigconf,screen]{acmart}
\AtBeginDocument{%
  }

\setcopyright{acmlicensed}
\copyrightyear{2026}
\acmYear{2026}
\setcopyright{cc}
\setcctype{by}
\acmConference[ICSE '26]{2026 IEEE/ACM 48th International Conference on Software Engineering}{April 12--18, 2026}{Rio de Janeiro, Brazil}
\acmBooktitle{2026 IEEE/ACM 48th International Conference on Software Engineering (ICSE '26), April 12--18, 2026, Rio de Janeiro, Brazil}
\acmPrice{}
\acmDOI{10.1145/3744916.3773262}
\acmISBN{979-8-4007-2025-3/2026/04}

\usepackage{mystyle}

\begin{document}

\newcommand{\myauthornote}[3]{{\color{#2} {\sc #1}: #3}}
\newcommand{\hy}[1]{\myauthornote{HY}{red}{#1}}
\newcommand{\cy}[1]{\textcolor{blue}{\textbf{Chen}: #1}}
\newcommand{\gu}[1]{\myauthornote{Gu}{blue}{#1}}

\newcommand{\diff}[1]{\textcolor{black}{#1}}

\title[A Semantic-based Optimization Approach for Repairing LLMs: Case Study on Code Generation]{A Semantic-based Optimization Approach for\\Repairing LLMs: Case Study on Code Generation}
\author{Jian Gu}
\email{jian.gu@monash.edu}
\affiliation{%
  \institution{Monash University}
  \city{Melbourne}
  \country{Australia}
}

\author{Aldeida Aleti}
\email{aldeida.aleti@monash.edu}
\affiliation{%
  \institution{Monash University}
  \city{Melbourne}
  \country{Australia}
}

\author{Chunyang Chen}
\email{chun-yang.chen@tum.de}
\affiliation{%
  \institution{Technical University of Munich}
  \city{Heilbronn}
  \country{Germany}
}
\additionalaffiliation{%
  \institution{Monash University}
  \city{Melbourne}
  \country{Australia}
}

\author{Hongyu Zhang}
\email{hyzhang@cqu.edu.cn}
\affiliation{%
  \institution{Chongqing University}
  \city{Chongqing}
  \country{China}
}

\begin{abstract}
Language Models (LMs) are widely used in software engineering for code generation, but they may produce erroneous code. Rather than repairing outputs, a more thorough remedy is to address underlying model failures. LM repair offers a lightweight solution: it requires minimal data, lowers computational cost, and limits side effects. Unlike full retraining, LM repair focuses on applying tailored updates to targeted neurons, making it suitable for limited resources, high-performance demands, or strict safety requirements.
In this paper, we propose \ul{S}emantic \ul{T}argeting for \ul{A}nalytical \ul{R}epair (\textsc{STAR}), a novel semantic-based optimization method for repairing LLMs.
\textsc{STAR} realizes the main operations of repairing LMs in an optimization process, including locating ``buggy neurons'', solving ``neuron patches'', and patching ``buggy neurons''.
The neuron patches are computed with a solid semantic-based analytical formula, which directly bridges the changes to logits with the deltas of neurons, by steering latent representations.
Compared to the prior work of LM repair (\textsc{MINT}) and standard optimization methods (\textsc{SGD}), \textsc{STAR} integrates their strengths while mitigating their limitations.
By reformulating LM repair as an optimization process, \textsc{STAR} may solve multiple failures together, significantly improving the usefulness.
Evaluated on coding tasks using popular code LMs, \textsc{STAR} demonstrates superior effectiveness compared with the state-of-the-art.
For instance, on CoNaLa, \textsc{STAR} obtained 10.5\%--19.9\% improvements in ExactMatch score.
Besides, \textsc{STAR} exhibits better efficiency, achieving an average of 2.4--7.0 times speedup in solving each model failure.
In terms of side effects, namely the balance between generalization and specificity, \textsc{STAR} outperforms prior work by a significant margin.
Additionally, we conducted assessments on the overfitting risk of LM repair as well as the cumulative impact.
Further, we analyzed the differences with pipeline-based methods and explained the reason why \textsc{STAR} is better and how it mitigated the common limitations of LM repair.
\end{abstract}



\maketitle

\section{Introduction}
\label{sec:introduction}
As language models (LMs), particularly code-focused LMs, demonstrate remarkable potential in productivity activities, such as code generation and assisted programming~\cite{gu2023towards,Jiang2024ASO,Lyu2024AutomaticPL}, their limitations tend to grow more pronounced over time~\cite{10.1145/3709353}.
These models may become obsolete, produce unsafe outputs, and fail to incorporate critical updates, such as dependency changes or framework revisions~\cite{Dong2022CalibratingFK,Jayasuriya2023UnderstandingBC}.
These changes result in vulnerabilities and errors that cascade into downstream consequences, including security risks or operational failures.
For instance, code LMs producing buggy or insecure programs may introduce systemic risks~\cite{Li2024BadEditBL,Li2024BackdoorLLMAC}, while even subtle inaccuracies, like flawed numerical reasoning (such as wrongly stating $9.8 < 9.11$), may propagate into critical failures~\cite{web_transluce_20241031}.
Rather than post-processing LM outputs, a more principled way to rectify flaws is to address the underlying model failures.

Common approaches like rule-based methods and memory-based methods maintain a rule base or a memory store for the changes, which can be represented as plain text, hidden states, etc, but they prove inadequate.
Rule-based methods lack flexibility and are separate from models, so they fail to utilize the interconnected, high-dimensional nature of LMs~\cite{Yao2023EditingLL,Kirkpatrick2016OvercomingCF}. Memory-based methods typically require frequent training of an additional classifier as the memory expands, and show the risks of dimensional collapse, which undermines the reliability~\cite{Das2024LarimarLL,Dohmatob2024StrongMC}.
This underscores the necessity of \emph{language model repair (LM repair)}, which directly manipulates model parameters to solve model failures.

LM repair is an important and emergent topic in optimizing LMs.
It indicates lightweight solutions with targeting and tailoring updates, while staying lightweight in computational costs and data demands.
LM repair resembles the topic of neural network repair (NN repair), however, NN repair methods mainly study traditional RNNs and CNNs~\cite{Sun2022CausalityBasedNN,LiCalsi2023AdaptiveSR,Xie2021RNNRepairAR,Ma2024VereVG}.
Due to the differences in model structure, mechanism of model functionality, as well as model scales, they cannot serve for repairing LMs.
Meanwhile, LM repair differs from other techniques that directly modify model parameters, including model training and model editing.
Model training methods fail to maintain the stability of LM behaviors, such as keeping the pattern of neuron firing as stable as possible~\cite{Du2018TechniquesFI}.
Therefore, they demand extensive data resources to avoid side effects, such as catastrophic forgetting~\cite{Kirkpatrick2016OvercomingCF}.
% ...
Model editing methods mainly focus on knowledge-related tasks, instead of general next-token prediction tasks. Also, they require an additional corpus to cover as much knowledge as possible, and an inference process to cache massive intermediate results~\cite{He2025KnowledgeUN}, so lack the feasibility of repairing LMs.

The prior work of LM repair is a pioneering pipeline but faces several limitations:
First, it is not guaranteed that model failures will be solved. This is caused by heuristic decisions in repairing LMs. For example, \textsc{MINT}~\cite{gu2023neuron} solves LM failure by patching merely one neuron. Its underlying hypothesis is that neurons and knowledge are one-to-one associations, which remains controversial~\cite{wei2024does}.
%
Second, the computation cost of LM repair may be very expensive. Lots of computational time overhead comes from: the repeated computations of forward-pass and backward-pass; the sequential operations of updating multiple neurons; and the sequential process of solving multiple failures.
Third, there is still room for further improvements regarding side effects. Due to the black-box nature of neural models, updating any model parameters will inevitably impact the initial capability of LMs and any earlier updates. It is hard to know which parameters to update and how to update them.

To mitigate the limitations, the countermeasure is to \emph{formulate LM repair as an optimization process}.
Compared to existing pipeline methods, an optimization method shows multiple advantages.
For example, it reduces the repeated computations and updates multiple neurons together. That means a significantly better efficiency.
Also, it guarantees reduced side effects compared to existing optimization methods, by selectively updating neurons~\cite{fu2023effectiveness}.
Moreover, it turns to an alternative hypothesis suggesting that neurons and knowledge follow many-to-many dynamic associations~\cite{allen-zhu2025physics}, which is consistent with recent empirical findings~\cite{conmy2023towards,chen2025identifying}.
The research gap is, \emph{how to realize locating and patching certain neurons in optimization}, and this is exactly the focus of this paper.

We propose a novel optimization method for LM repair: \ul{S}emantic \ul{T}argeting for \ul{A}nalytical \ul{R}epair, shortened as \textbf{\textsc{STAR}}.
\textsc{STAR} operates on the granularity of neurons, and we call the neuron that is critical to certain failures and meanwhile beneficial in solving failures a ``buggy neuron''.
The mechanism of \textsc{STAR} contains three operations: i) locating ``buggy neurons''; ii) patching ``buggy neurons''; iii) solving ``neuron patches''.
Intuitively, they correspond to the concepts of locating and patching buggy codelines in program repair, as well as solving code patches.
They rely on the same intermediate results of LMs, so they are combined together in an optimization process.
\textsc{STAR} is fast and useful in repairing LMs. It can solve model failures with few exemplary data, by updating one or two neurons, without causing side effects.
In this paper, we mainly focus on the effects of LM repair on code generation as a case study.
We conduct experiments on three coding datasets, with five modern and powerful code LMs.
The results show that \textsc{STAR} significantly outperforms the state-of-the-art (SOTA) in terms of effectiveness and efficiency, and show marginal side effects.
Also, the statistical insights show that the sparsity pattern matters in LM repair, and a proper selection of layers and neurons to update decides the effects of LM repair.
In terms of side effects, our approach shows a significantly better balance between generalization and specificity.

\noindent
To summarize, the contributions of this paper are as follows:
\begin{itemize}
    \item We formulated LM repair as an optimization process, and implemented an optimization method, which is more effective and efficient than the SOTA, with reduced side effects.
    The replication artifact is available online for open science \footnote{\url{https://github.com/jianguda/star}}.
    \item We proposed a semantic-based analytical framework for solving the updates to knowledge neurons. When using it as the prior guidance, the optimization has a fast convergence.
    \item We conducted extensive experiments reporting the empirical insights on the effects of sparsity pattern on LM repair. It showcases the necessity of layer selection and improves robustness in neuron targeting.
\end{itemize}

\section{Preliminaries}
\label{sec:preliminaries}

\subsection{LM Failures in Next-token Prediction}

For a given language model that performs next-token prediction, including coding tasks performed by code LMs, the ground truth is known as the \textit{target token}, and the token having the largest probability is known as \textit{argmax token}.
The two tokens shall be identical; if not, it indicates a model failure.
From the viewpoint of next-token prediction, LM repair methods solve model failures by either increasing the probability of the target token or decreasing the probability of the argmax token, until they become the same.

\subsection{LM Semantics in Latent Space}

\begin{figure}[htbp]
    \centering
    \includegraphics[width=\linewidth]{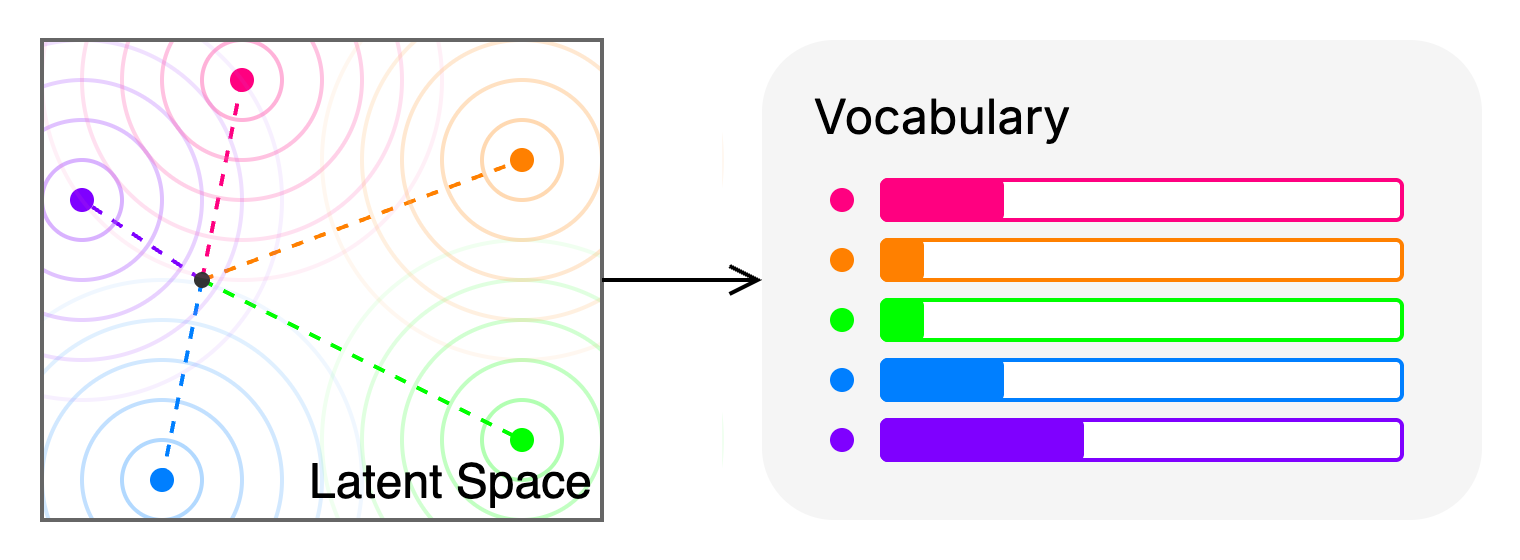}
    \caption{Semantics-related concepts in LM latent space.}
    \label{fig:semantic}
\end{figure}

Vocabulary-defined semantics is a semantic theory for language models~\cite{gu2024vds}. For the recognizable semantic meanings of a given LM, it proposed defining a set of special representations in the latent space to associate with the labels in the vocabulary.
Using these defined special representations, it is possible to compute logits directly in LM latent space, instead of on LM vocabulary~\cite{gu2024vds,gu2025salf}.
The semantics-related concepts are illustrated in \cref{fig:semantic}, which will be introduced separately later. For convenience, we call the process of solving them ``semantic solving''.

\paragraph{Semantic Basis}
A label in LM vocabulary is one symbol of each typical semantic meaning that LM can understand or express.
Each label has an associated representation in LM latent space, termed as ``semantic basis''~\cite{gu2024vds}.
Semantic bases accurately represent the label space of model outputs and comprehensively encompass all output labels within the vocabulary.
They are the semantic equivalences of vocabulary labels in latent space.
Semantic bases obtained by computing the pseudoinverse of the LM-head matrix, incurring a relatively low computational cost.
In \cref{fig:semantic}, semantic bases are the color dots in latent space, corresponding to vocabulary labels.

\paragraph{Semantic Logits}
The complex semantic meaning of any latent representation is quantified by measuring its distances to all semantic bases, termed as ``semantic logits''~\cite{gu2024vds}.
In numerical terms, the logits calculated by distance measurement is equivalent to the common calculation by matrix multiplication.
However, semantic logits show great advantages in amplifying or suppressing a certain semantic meaning.
For an arbitrary representation, illustrated as the dark dot in \cref{fig:semantic}, by steering it to the direction of a certain semantic basis, the probability of the corresponding vocabulary label will increase, and the label becomes more likely to be predicted. The semantic logits is illustrated as the dashed lines between the dark dot and the color dots.
In addition, semantic logits can be computed normally in the latent space of any model layer, since the semantic property of LM latent space remains stable across layers~\cite{gu2025beyond,gu2025seme}.

\section{Approach}
\label{sec:approach}



\begin{figure}[!ht]
    \centering
    \includegraphics[width=1.0\linewidth]{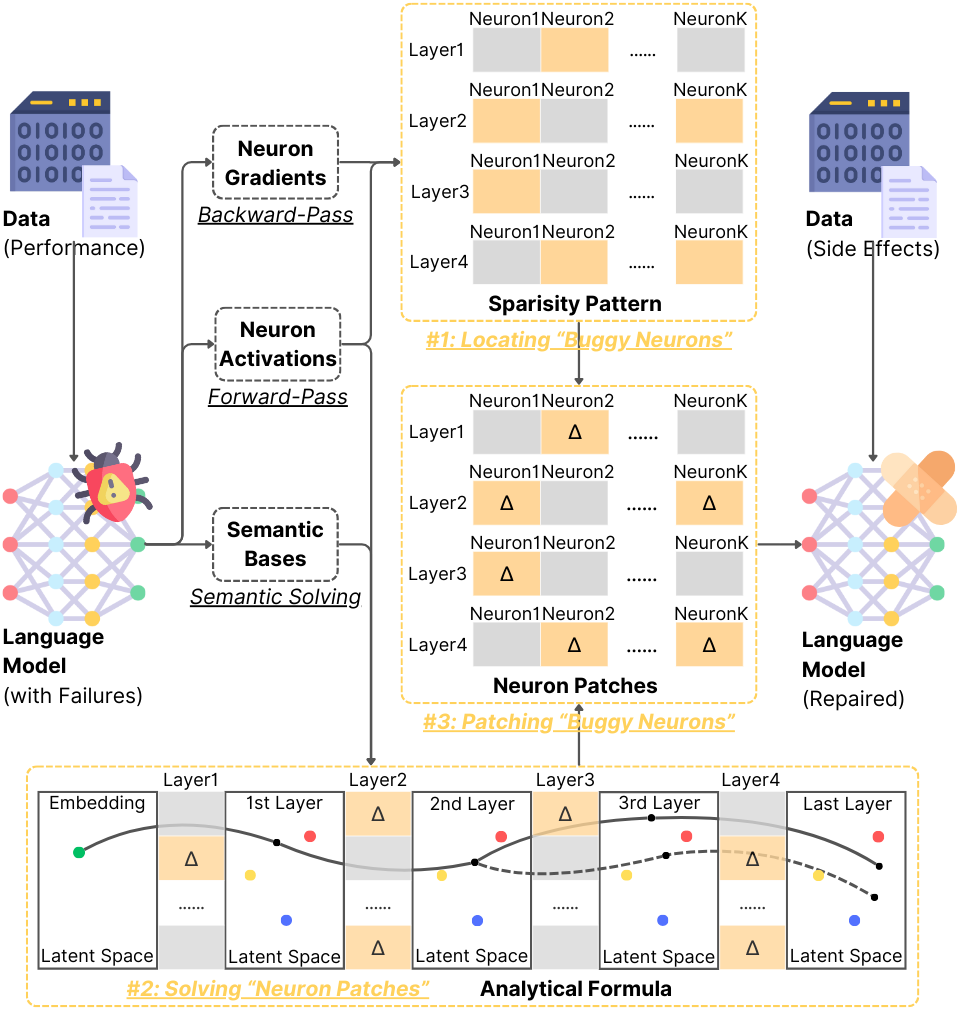}
    \caption{Overview of our approach for model repair.}
    \label{fig:overview}
\end{figure}

Our approach \textsc{STAR} is an optimization process, leveraging semantic and structural analysis of LM behaviors. It follows a paradigm of program repair but focuses on LMs instead.
% ...
In repairing LM failures, \textsc{STAR} operates on the granularity of neurons. A neuron means a vector in the weight matrix. We call the neuron that is critical to certain failures and meanwhile beneficial in solving failures as ``buggy neuron''.
As a 4-layer LM described in \cref{fig:overview}, we first conduct a forward-pass and a backward-pass to obtain the activations and gradients, and also, do semantic solving to obtain the semantic bases. Then, in the optimization process, we do the following operations:
(1) locating ``buggy neurons'': we employ attribution methods to specify the scope of neurons to update, to maintain the LM stability;
(2) solving ``neuron patches'': we use a novel analytical framework to solve the updates to any neuron, which we refer to as ``patch'';
(3) patching ``buggy neurons'': we integrate the neuron patch as the prior guidance in LM optimization, to guarantee gradual improvements.
In \cref{fig:overview}, we mark buggy neurons with the triangle symbol, and analyze layer-wise semantics with curves and dots.
Intuitively, the operations of locating and patching buggy codelines in program repair, as well as solving code patches. We elaborate on them in the following subsections.

\subsection{Locating Buggy Neurons}

The operation of locating ``buggy neurons'' is similar to the operation of locating buggy codelines in program repair. It requires attribution methods to locate buggy neurons and notable layers.




Our approach \textsc{STAR} has underlying supportive evidence from the related study on LM interpretability:
First, \textsc{STAR} operates at the neuron-level granularity, not parameter-level.
It has been proven that certain neurons in LMs contribute more than others to the probability of a given output~\cite{Dhamdhere2018HowII}.
For example, neuron firing indicates the process of individual and groups of model neurons spiking (response with significant activations) in response to the given outputs~\cite{Du2018TechniquesFI}.
Moreover, \textsc{STAR} updates multiple neurons in solving a model failure. Based on the hypothesis that neurons and knowledge follow many-to-many dynamic associations~\cite{allen-zhu2025physics}, instead of controversial one-to-one associations~\cite{wei2024does}.
Last, \textsc{STAR} operates on the hidden layer of feed-forward networks (FFN), in which neurons are defined as \textit{``knowledge neurons''}~\cite{Dai2021KnowledgeNI}. Knowledge neurons are positively correlated to the knowledge expression of LMs. \textsc{STAR} introduces no changes to multi-head attention (MHA).

We focus on the neurons critical in causing a model failure and beneficial in solving it, and we term them as ``buggy neurons''.
For a model failure, if a neuron has high activation scores in LM forward-pass, it is critical in causing the failure; and if it also has high gradient scores in LM backward-pass, it is beneficial in solving the failure. Therefore, we use both activations and gradients to locate the buggy neurons~\cite{Zhou2021DoFA}.
Based on the related work in deep learning interpretability~\cite{Deng2023UnderstandingAU}, \textit{feature attribution} algorithms are exactly ready-to-use methods to help locate buggy neurons~\cite{Chattopadhyay2019NeuralNA}.

For a given model failure, we employ the \textit{Input X Gradient}~\cite{Ancona2017AUV} algorithm to locate the buggy neurons.
For a given token $t$, \textit{Input X Gradient} multiplies its gradient with the input embedding and then takes the L2 normalization of the resulting vector. The computation of the attribution score is formalized as: ${score} = \left\|\nabla_{X_i} f_t\left(X_{1: n}\right) X_i\right\|_2$, where $X_i$ is the input embedding at step i, $\nabla_{X_i} f_t\left(X_{1: n}\right)$ is the gradient of token $t$.
The method assigns an attribution score to each neuron, which is the element-wise product of the input and the gradients.
The first term represents the contribution of a neuron in causing the failure, while the second term represents the benefit of solving the failure by updating that neuron.
The attribution score approximates the extent to which a neuron is buggy. A higher attribution score means a larger likelihood of being buggy.

We emphasize reducing the side effects of LM repair by selectively updating buggy neurons, or neuron targeting.
The side effects indicate the unexpected changes to the capability of LM concerning the generalization and specificity.
In the context of LM repair, \emph{generalization} refers to the repaired model's ability to perform well with new, previously unseen data, drawn from the same distribution as the one used to edit the model;
while \emph{specificity} refers to the ability of the LM repair technique to apply changes that only impact the target tokens and not the other unrelated tokens. In other words, it reveals the ability of LM repair to correctly identify unrelated tokens and thus minimize the impact on those tokens.




When locating buggy neurons, we additionally take into consideration the effects of the hierarchical structure of LMs.
On one hand, it aligns with well-established findings that lower layers capture shallow patterns while upper layers capture deep semantics~\cite{Lu2019UnderstandingAI,Rogers2020API}.
On the other hand, caused by normalization and non-linear activation, the gradients across layers undergo numerical instability, making the comparison between the attribution scores from different layers unfair.
There are two types of sparsity patterns of buggy neurons.
The common practice is locating buggy neurons by doing attribution directly on the model level, labeled as \expttag{[model-wise]}. Besides, we suggest first selecting one or more layers based on their importance to model performance, then locating buggy neurons within the scope, labeled as \expttag{[layer-wise]}.

Similar to using attribution methods to locate buggy neurons, we use an occlusion-based method to select notable layers. It generalizes the idea of feature occlusion~\cite{Zintgraf2017VisualizingDN}, which interprets the decision of neural networks by emphasizing the removal of input features and measuring impact. Instead, we respectively disable each model layer and observe the changes to the model loss. The magnitude of loss change is an approximation measurement of a layer being the notable layer. A larger change in loss indicates a higher likelihood of being more notable in LM repair.

The amount of buggy neurons and notable layers play an important role in reducing the side effects while guaranteeing the effects of LM repair. To the best of our knowledge, there is no reliable conclusion from the prior work showing the empirical selections. We conduct extensive experiments to form statistical insights.

\subsection{Solving Neuron Patches}

The operation of solving ``neuron patches'' is similar to the operation of solving code patches in program repair. We address this challenge from a novel perspective of LM optimization by leveraging the semantic property of the LM latent space (refer to \cref{sec:preliminaries}).

In next-token prediction, the last token of the inputs decides the initial latent representation of the output token~\cite{gu2025salf}.
Our approach propose steering the latent representation to the ground truth.
The effects of LM repair to latent representations of the last token are illustrated with a 4-layer LM, as shown \cref{fig:overview}.
The vocabulary is a collection of colorful tokens, and the corresponding semantic bases are shown as the color dots.
The dark curves represent the transition trace of the output token, defined by the latent representations in each latent space (the dark dots).
Starting from the input-side semantic basis (the green dot), the latent representation undergoes a gradual transition (the solid dark curve). The dark dot in the last-layer latent space is close to the red semantic basis, so the argmax token is \emph{red}.
However, the ground truth is blue, so we conduct LM repair by patching the buggy neurons in the 3rd layer, causing a new trace (the dashed dark curve). The dark dot in the last-layer latent space becomes close enough to the blue semantic basis, so the argmax token becomes \emph{blue}.

\paragraph{Computation of Semantic Bases}
The logits at the output side is computed via matrix multiplications, that is $\vec{r}\cdot\mathbb{W}=\vec{l}$ where $\vec{r}$ is the resulting representation of the last model layer.
Since the loss is computed using the logits and the given ground truth. For each token to generate, the ground truth is used as its onehot embedding (for example, in cross-entropy loss), denoted with $\vec{e}$.
Therefore, we conduct a reversed computation using the pseudoinverse of LM-head matrix $\mathbb{W}_{o}^+$ to obtain the output-side semantic basis $\vec{s}_{o}=\vec{e}_{o}\cdot\mathbb{W}_{o}^+$.
% ...
We do a similar operation at the input side, to compute the semantic basis of each input token. That is, multiplying its onehot embedding $\vec{e}_{i}$ by the embedding matrix $\mathbb{W}_{i}$ to obtain the input-side semantic basis $\vec{s}_{i}=\vec{e}_{i}\cdot\mathbb{W}_{i}$.
It is important to highlight that, $\mathbb{W}_{i}$ and $\mathbb{W}_{o}^+$ are different in values even though their shapes are the same, so for the same token, the corresponding input-side semantic basis and output-side semantic basis are very different.

Our observation is that \textit{by updating the parameters of buggy neurons in a certain layer, the latent representation will be steered towards the semantic basis of the corresponding ground truth, therefore improving the logits}.
The initial parameters, the current representation, and the steering destination are known, while the updated parameters are unknown which is what we are interested in.
Considering the inputs to model parameters are known as well, we simplify our observation as a linear inverse problem, which is a typical optimization problem in the form of $\vec{a}\cdot\mathbb{X}=\vec{b}$, where $\vec{a}$ and $\vec{b}$ are known, and the objective is solving $\mathbb{X}$.
We propose a novel analytical framework that derives parameter updates by tracing the changes from logits and through representations, namely the causal chain: \emph{$\Delta{logits} \rightarrow \Delta{representations} \rightarrow \Delta{parameters}$}.

Based on the intended logits changes, we first compute the deltas in representations, and then solve the deltas in parameters.
Leveraging vocabulary-defined semantics, we know on-the-fly how good a latent representation is concerning the ground truth by measuring its distance to the corresponding semantic basis.
For computational simplicity, we measure the (normalized) differences between the semantic bases of the argmax token and that of the target token, noted as $\vec{s}_{\Delta}$. The equation is: $\vec{s}_{\Delta} = \texttt{norm}\left(\vec{s}_{o\left(\texttt{argmax}\right)} - ~\vec{s}_{o\left(\texttt{target}\right)}\right)$.





Based on the geometric meaning of vector addition, adding the semantic basis to the neuron weight in the FFN output layer will steer the latent representation to the semantic basis.
Let's say $\vec{v}$ is the intermediate result in the FFN module (the outputs of the first weight matrix and the inputs of the second weight matrix).
The computation equation is: $\vec{v} \cdot W_{\Delta} = ~\vec{s}_{\Delta}$.



To solve for the weight matrix $W_{\Delta}$ given input $\vec{v}$ and target $\vec{s}_{\Delta}$, we use a linear least squares solver. This function minimizes the Frobenius norm of the residual, effectively solving the optimization $\min_{W_{\Delta}} \| \vec{v} \cdot W_{\Delta} - \vec{s}_{\Delta} \|_F^2$ in the least squares sense. The resulting $W_{\Delta}$ has dimensions that allow it to map from the input space of $\vec{v}$ to the output space of $\vec{s}_{\Delta}$.
In updating model parameters, the deltas $W_{\Delta}$ will be accompanied with a coefficient factor $\alpha$, to control the efficacy and potential side effects: $W^\prime = W + \alpha~W_{\Delta}$.

From the process of solving neuron patches, we see that the minimal granularity for LM repair shall be neuron-level, since the changes in a neuron guarantee steering the latent representation to a meaningful optimization direction.
In contrast, parameter-level granularity taken by finetuning methods lacks such guarantee.







\subsection{Patching Buggy Neurons}

The operation of patching ``buggy neurons'' is similar to the operation of patching buggy codelines in program repair. The difference is that, in repairing LMs, the way to patch neurons will significantly affect the efficiency; while for program repair, it is very direct, simply overwriting the buggy codelines.

The common practice for parameter optimization is gradient descent methods, where model parameters are updated directly with the gradients.
However, the core strategy of gradient-based methods is merely exploring the loss surface. Since the location of the global optimum is unknown, they reduce the loss in a greedy style, following the steepest gradient direction.

To speed up parameter optimization, we introduce a novel process of \emph{shortest-path planning} to complement the general process \emph{loss-surface exploration} in our approach \textsc{STAR}.
Their differences lie in that, shortest-path planning requires the prior information of the global optimal (namely the ground truth) to guarantee the fastest convergence, while loss-surface exploration requires the gradients to guarantee the most greedy loss reduction.
In detail, the prior information to be introduced for shortest-path planning is the semantic delta of the weight matrix, namely the results of the semantic-based analytical framework.


\begin{algorithm}
\caption{Gradient Descent with Prior Guidance.}
\label{algo:optimizer}
\begin{algorithmic}[1]

\REQUIRE epoch number $e$; learning rate $\eta$; weight-matrix gradients $G$; weight-matrix deltas $W_{\Delta}$
\ENSURE weight-matrix parameters $W$
    \FOR{$t = 1$ to $e$}
        \STATE \textit{// element-wise sign compare}
        \STATE $f_t \gets \text{sign}(G_t) \odot \text{sign}(W_{\Delta})$
        \STATE \textit{// element-wise product}
        \STATE $u_t \gets W_{\Delta} \odot f_t$
        \STATE \textit{// element-wise update}
        \STATE $W_t \gets W_{t-1} - \eta u_t$
    \ENDFOR
\end{algorithmic}
\end{algorithm}

We combine the two processes, using the sign information of gradients and the magnitude information of priors.
%
The key insight here is that the gradient sign provides directional information, while the prior’s magnitude acts as an adaptive scaling factor that modulates step sizes based on the location of the global optimum.
%
On one hand, the gradient sign ensures updates are moving in the correct direction to minimize loss; On the other hand, the prior magnitude guarantees the location of the global optimum is approaching.
In contrast, when doing parameter optimization using only the gradient sign, the updates are uniform in size regardless of how much confidence we have in different directions; using only the prior’s magnitude can lead to suboptimal updates since it lacks real-time feedback from the loss.
As described in \cref{algo:optimizer}, our approach \textsc{STAR} decides the model parameters to update based on the signs from gradients and the magnitudes from priors (the weight matrix deltas).
This combination creates a mechanism of parameter optimization that is magnitude-aware (from the prior) and correct in direction (from the gradient), leading to faster and more stable optimization.

\begin{figure}[!ht]
    \centering
    \includegraphics[width=\linewidth]{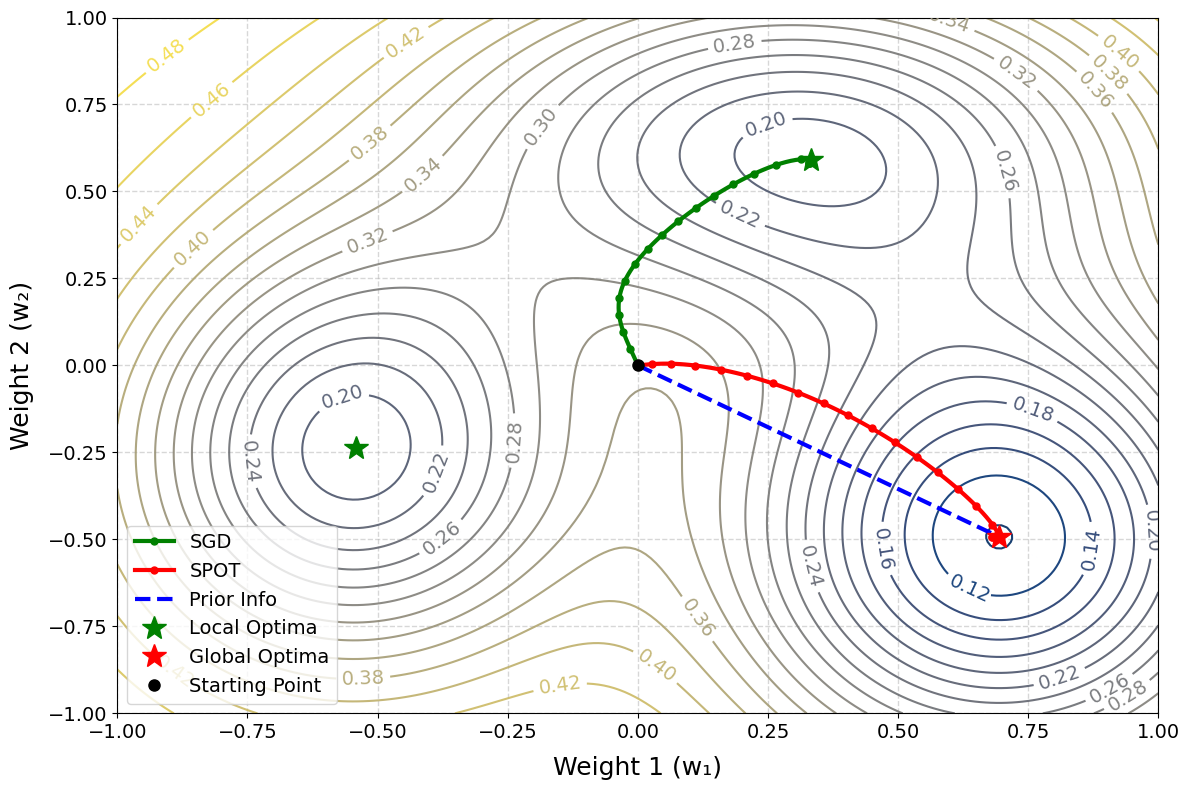}
    \caption{Illustration of \textsc{SGD} and \textsc{STAR} in the loss surface during parameter optimization.}
    \label{fig:loss-surface}
\end{figure}

We illustrate the difference between \textsc{SGD} with \textsc{STAR} in the loss surface of an actual parameter optimization, a surface about parameter combination (the coordinate locations) and corresponding loss (the contour lines), where the step size is $1e-2$ and the step number is $100$.
As shown in \cref{fig:loss-surface}, \textsc{SGD} and \textsc{STAR} differ in the converged locations. Green stars indicate the local optima, and the red star indicates the global optima.
The updates start from the origin of the coordinate system, meaning the initial parameters are 0s (the loss is around $0.33$).
Using merely the gradients, the updates follow a green curve (the \textsc{SGD} trajectory) approaching the local optima (the loss is around $0.18$). In contrast, with the prior information (the dashed line), the updates follow a red curve (the \textsc{STAR} trajectory) approaching the global optima (the loss is around $0.09$).




\section{Experiments}
\label{sec:experiments}

We formulate our research questions and introduce the experimental setups, including baselines, models, datasets, and metrics.
We prioritize following the choices of previous work on LM repair~\cite{gu2023neuron}.


\begin{reqs}

\item [\req{1}] \textbf{How effective and efficient is \textsc{STAR} in repairing LMs?}

\end{reqs}




\paragraph{Setup}
We conduct experiments to compare \textsc{STAR} with baselines on code generation tasks.
For each data used in repairing LM, we use the same data to validate that the model failure is indeed solved.
Besides the working mode of solving failures one-by-one in succession (marked as \textit{single}), \textsc{STAR} supports solving failures in a small batch together (marked as \textit{multiple}). We evaluate \textsc{STAR} in both working modes, and compare it with the corresponding baselines.

\paragraph{Measures}
We measure the effectiveness by comparing the quality of code generation before and after LM repair, using the following metrics (larger values are better):
\textit{Exact Match} means the proportion of generated tokens that perfectly match the ground truth;
\textit{BLEU}~\cite{Papineni2002BleuAM} computes the average proportion of overlapping $n$-gram (typically \num{4}-gram) between the prediction and its ground truth;
% ...
Meanwhile, we measure the efficiency of LM repair based on the elapsed execution time in solving each model failure, denoted as \textit{Time Cost} (smaller values are better).






\paragraph{Overfitting Risk}
Repairing LMs using a single data each time may cause the overfitting concern. Therefore, we conduct an additional hold-out test to assess the overfitting risk.
In detail, for a given dataset, we use $80\%$ data for LM repair and the left $20\%$ as the hold-out set. The LM performance (based on BLEU score) on the hold-out set before any repairs, marked $P_\texttt{base}$. Each time we do LM repair using the $k$-th data, we measure the updated LM performance on the hold-out set, marked $P_\texttt{k}$, and then compute \textit{Relative Performance Drop (RPD)}, defined as $\mathtt{RPD_\texttt{k}} = (P_\texttt{base} - P_\texttt{k}) / P_\texttt{base}$. This aligns with common overfitting measures in model finetuning~\cite{finn2017model,Dodge2020FineTuningPL}.
We compute statistical values of the computed RPD scores to analyze the overfitting risk: \textit{Mean}, \textit{Std}, and \textit{Inter-Quartile Range (IQR)}. \texttt{IQR} means the difference between the 75th percentile (Q3) and the 25th percentile (Q1), so avoids over-weighting outliers. The smaller, the better. Meanwhile, we measure the percentage of degradation cases. A smaller percentage means it is less likely to be overfitting.

\begin{reqs}

\item [\req{2}] \textbf{How well does \textsc{STAR} perform with varying sparsity patterns in repairing LMs?}

\end{reqs}





\paragraph{Setup}
The experimental design is the same as RQ1. However, we measure the effects of sparsity patterns.
It means that, when repairing LM failures, we are using different variants of \textsc{STAR}.
First, we compare the model-wise variant with the layer-wise variant when given a limited amount of neurons. Depending on the used LM, we take the amount of neurons in a layer and gradually reduce it proportionally.
Then, we study layer-level sparsity patterns and neuron-level sparsity patterns. For the former, we select a different number of model layers by certain proportions; for the latter, we locate a certain amount of neurons in a certain model layer.


\paragraph{Measures}
To measure the effects of varying sparsity patterns on the performance of LM repair, we compute and compare the proportions of solved LM failures, namely \textit{accuracy}.
It highlights the differences between the \textsc{STAR} variants with similar configurations.


\begin{reqs}

\item [\req{3}] \textbf{What are the side effects of STAR in repairing LMs?}

\end{reqs}



\paragraph{Setup}
The experiments are carried out on a benchmark specifically designed for side effects~\cite{gu2023neuron}.
For each data used in repairing LM, we use other data to measure the side effects, concerning generalization and specificity.
For generalization, we use the related data sharing the same ground truth to study generalization; for specificity, we use the unrelated data having different ground truth to study specificity.

\paragraph{Measures}
Given that the goal of model repair is to elevate the target token to the argmax token, we measure the change to the probability gap between the target token and the argmax token, where $gap = p_\texttt{target} - p_\texttt{argmax}$, and the expected gap $\hat{gap}$ is the initial value before repairing.
The magnitude of changes in token probabilities shows broader trends in the model's behavior, regardless of whether the changes lead to actual changes in model performance (such as accuracy).
We compute \textit{Mean Absolute Error (MAE)} and \textit{Root Mean Square Error (RMSE)} of the gap changes on related data for generalization, and on unrelated data for specificity.
% ...
Let $\texttt{G}$ denote the generalization metric, measuring the positive effects on the related data, and $\texttt{S}$ denote the specificity metric, measuring the negative effects on the unrelated data. For both $\texttt{G}$ and $\texttt{S}$, a large score means a better effect.
% ...
Although generalization and specificity are orthogonal in concepts, they often exhibit a trade-off in practice. We quantify the balance between them using their harmonic-mean value, namely \textit{Generalization–Specificity Harmonic (GSH)}, defined as $\texttt{GSH} = \frac{2 \cdot\texttt{G} \cdot \texttt{S}}{\texttt{G} + \texttt{S}}$.
The harmonic formulation ensures that GSH score penalizes repairing methods with imbalanced performance, where either generalization or specificity is poor.
A larger GSH score indicates a better balance between generalization and specificity.

\paragraph{Cumulative Impact}
Due to the inherent complexity and black-box nature of neural models, LM repair methods are hard to be free of side effects, even though they already perform well in fixing immediate model failures. We therefore study the model's performance stability after heavy repairs, where potential side effects tend to accumulate and become apparent.
In detail, we conduct LM repair on long code generation. When the ground truth is long enough, LMs have to handle more failures and undergo multiple rounds of repair. The evaluation metrics are ExactMatch and BLEU.

\subsection{Baselines}

We mainly compare our approach \textsc{STAR} with the state-of-the-art of LM repair \textsc{MINT}~\cite{gu2023neuron}, following the default settings as recommended.
Besides, we introduce the gradient descent method using \textsc{SGD} optimizer as an additional baseline, labeled as \textsc{SGD}, since it means the most general practice in finetuning LMs.
The scope of parameters to optimize is constrained be the second weight matrix of FFN modules, for fair comparisons with \textsc{MINT}.



\subsection{Models}
\label{subsec:model}

The code language models used in our experiments include \textsc{CodeLlama} (7B)~\cite{Rozire2023CodeLO}, \textsc{StarCoder2} (7B)~\cite{Lozhkov2024StarCoder2A}, \textsc{OpenCoder} (8B)~\cite{Huang2024OpenCoderTO} and \textsc{Qwen2.5-Coder} (7B, 14B)~\cite{Hui2024Qwen25CoderTR}.
They are competitive and modern open-weight LMs for coding tasks.
The model details are shown in \cref{tab:model}.
In the experiments, Qwen2.5-Coder models are loaded in BF16 to reduce both memory and computation costs. The impact on model performance shall be very marginal as the model parameters are released in BF16~\cite{Micikevicius2017MixedPT}. For the sparsity pattern study in RQ2, we use the LMs that normally load in single-precision, since their number of parameters is very close.

\begin{table}[!tb]
    \caption{Details of Code Language Models}
    \label{tab:model}
    \centering
    \resizebox{\linewidth}{!}{
        \sisetup{table-format=6}
\begin{tabular}{
    l ccc cc
}

\toprule

& \multicolumn{1}{c}{\textbf{CodeLlama}} & \multicolumn{1}{c}{\textbf{StarCoder2}} & \multicolumn{1}{c}{\textbf{OpenCoder}} & \multicolumn{2}{c}{\diff{\textbf{Qwen2.5-Coder}}} \\
\cmidrule(lr){2-2} \cmidrule(lr){3-3} \cmidrule(lr){4-4} \cmidrule(lr){5-6}
& \textbf{7B} & \textbf{7B} & \textbf{8B} & \diff{\textbf{7B}} & \diff{\textbf{14B}} \\

\midrule
Release Date & {Aug. 2023} & {Feb. 2024} & {Nov. 2024} & \multicolumn{2}{c}{Sept. 2024} \\
\midrule
Corpus Scale & 2.5T tokens & 3.0T tokens & 2.5T tokens & \multicolumn{2}{c}{5.5T tokens} \\
\midrule
Layer Num. & 32 & 32 & 32 & 28 & 48 \\
\midrule
Dimension & 4,096 & 4,608 & 4,096 & 3,584 & 5,120 \\
\midrule
Vocabulary & 32,016 & 49,152 & 96,640 & \multicolumn{2}{c}{151,646} \\
\bottomrule

\end{tabular}

 }
\end{table}

\subsection{Datasets}
\label{subsec:dataset}





Following the settings of prior work, we study \req{1} and \req{2} with code generation datasets, and study \req{3} with an additional benchmark and code completion datasets. The statistics of datasets are shown in \cref{tab:corpus}. The benchmark is introduced separately.


\begin{table}[!tb]
    \caption{Statistics of Code Generation Datasets}
    \label{tab:corpus}
    \centering
    \resizebox{\linewidth}{!}{
        \sisetup{table-format=6}
\begin{tabular}{
    ll rrrrr
}

\toprule

& & \textbf{CoNaLa} & \textbf{IA32} & \textbf{TLDR} & \diff{\textbf{HumanEval}} & \diff{\textbf{BigCode}} \\





\midrule
\multirow{2}*{Amount}
& Retrieval & 2,379 & 2,560 & 6,414 & \textemdash & \textemdash \\
& Inference & 500 & 640 & 928 & 164 & 148 \\

\midrule
\multirow{2}*{Length}
& Inputs & 58.2 & 46.8 & 46.9 & 450.6 & 155.1 \\
& Outputs & 43.1 & 16.0 & 35.7 & 146.7 & 622.0 \\


\bottomrule

\end{tabular}

 }
\end{table}

For RQ1 and RQ2, we employ three datasets for different types of code generation:
(1) \textbf{CoNaLa}~\cite{Yin2018LearningTM} is for \textit{Line-level Code Generation}: given a natural language intent, the model generates the most suitable code. It consists of pairs of rewritten intents (similar to interline comments) and Python codelines;
(2) \textbf{IA32}~\cite{Liguori2021CanWG} is for \textit{Shellcode Generation}: given a natural language intent, the model generates the corresponding assembly code. It consists of pairs of natural language intents and assembly code (for Intel Architecture, 32-bit);
(3) \textbf{TLDR}~\cite{Zhou2022DocCoderGC} is for \textit{Intent-to-Bash Translation}: given a natural language intent, the model generates the corresponding bash command. It is composed of pairs of natural language intents and bash commands.
When evaluating data in the inference division, we interact with LMs with few-shot promptings~\cite{Dong2022ASF}. We retrieve demonstrations from the retrieval division as additional context.
In addition, we study the overfitting risk with CoNaLa in \req{1}.

For RQ3, we use an additional benchmark to evaluate the generalization and specificity of LM repair~\cite{gu2023neuron}.
It has a total size of 450 data samples, grouping them by the type of failures.
For each time of repair, it evaluates generalization on 9 samples from the same group (whose ground truth is the same as the current data); and evaluates specificity on 14 samples from different groups (whose ground truth is different from the current data).
For generalization, the effects on related data are positive so should be maximal; while for specificity, the effects on unrelated data are negative so should be minimal.
Besides, we study the cumulative impact of LM repair with \textbf{HumanEval}~\cite{Chen2021EvaluatingLL} and \textbf{BigCode}Bench (hard set)~\cite{Zhuo2024BigCodeBenchBC}. They are code completion datasets, providing comprehensive docstrings as inputs and codelines as ground truth.
Compared to other datasets, HumanEval features input data and output data with 9 and 5 times greater; and BigCode features input data and output data with an average length 3 and 20 times greater. A longer ground truth requires solving more LM failures.
Depending on the actual LM used, each correct completion in HumanEval and BigCode on average requires solving 5--20 and 30--50 failures, while for each code generation (CoNaLa, IA32, TLDR), LMs usually handle less than $10$ failures.
In our experiments, LMs directly complete the missing codelines for any given docstring. LMs have no additional context so will expose as many as failures possible.
In addition, we study the cumulative impact with code completion datasets in \req{3}.

\subsection{Implementation Details}
\label{sec:exp:details}

We constrain the range of parameter updates to within $[-0.1, +0.1]$ to guarantee fair comparisons. This prevents extreme parameter updates that may excessively affect LM performance, such as any single dominant update.
For \textsc{SGD}, the learning rate is the recommended value $1e-2$, and the maximal optimization step is $10$.
For \textsc{MINT}, the factor is $1e-2$ and the maximum times of neuron patching is $10$.
For \textsc{STAR}, the factor is $1e-2$ and the maximum optimization step is $10$.
Experiments were run on Nvidia A100.


\section{Results}
\label{sec:results}

In the \req{1} results, the optimal scores are highlighted in bold.
In the \req{3} results, the optimal balance between generalization and specificity is highlighted in bold. To facilitate the analysis, the results better than \textsc{STAR} are emphasized in grey color.


\subsection{\req{1} Results}
\label{subsec:rq1_results}

From the perspective of parameter optimization, \textsc{MINT} only updates a neuron each time (a neuron means a vector in the second FFN weight matrix), while \textsc{SGD} updates all neurons each time (all vectors in the second FFN weight matrix). In contrast, \textsc{STAR} updates neurons selectively, taking advantage of \textsc{MINT} and \textsc{SGD} while avoiding their disadvantages, and therefore performing better.


\begin{table}[!ht]
    \caption{Effectiveness of LM Repair in Coding Tasks}
    \label{tab:results_rq1_effects}
    \centering
    \resizebox{\linewidth}{!}{%
        \sisetup{table-format=2.2}
\begin{tabular}{
    ll rrr rrr
}

\toprule

\multirow{2}[2]{*}{\textbf{Mode}} & \multirow{2}[2]{*}{\textbf{Method}} & \multicolumn{3}{c}{\textbf{ExactMatch $\uparrow$}} & \multicolumn{3}{c}{\textbf{BLEU $\uparrow$}} \\
\cmidrule(lr){3-5} \cmidrule(lr){6-8}
&
& {\textbf{CoNaLa}} & {\textbf{IA32}} & {\textbf{TLDR}}
& {\textbf{CoNaLa}} & {\textbf{IA32}} & {\textbf{TLDR}}
\\

\midrule
\multicolumn{2}{l}{\textbf{\textsc{CodeLlama-7B}}}
& 0.766 & 0.820 & 0.557
& 0.662 & 0.689 & 0.445
\\
\cmidrule(lr){1-8}
\multirow{2}{*}{\textit{\shortstack{Single}}}
& \expttag{+~\textsc{MINT}}
& 0.825 & 0.858 & 0.657
& 0.741 & 0.755 & 0.559
\\
& \expttag{+~\textsc{STAR}}
& 0.946 & \textbf{0.914} & 0.920
& 0.882 & \textbf{0.833} & 0.810
\\
\cmidrule(lr){1-8}
\multirow{2}{*}{\textit{\shortstack{Multi\\ple}}}
& \expttag{+~\textsc{SGD}}
& 0.909 & 0.816 & 0.810
& 0.868 & 0.713 & 0.811
\\
& \expttag{+~\textsc{STAR}}
& \textbf{0.948} & 0.881 & \textbf{0.944}
& \textbf{0.904} & 0.738 & \textbf{0.887}
\\

\midrule
\multicolumn{2}{l}{\textbf{\textsc{StarCoder2-7B}}}
& 0.808 & 0.924 & 0.589
& 0.716 & 0.932 & 0.486
\\
\cmidrule(lr){1-8}
\multirow{2}{*}{\textit{\shortstack{Single}}}
& \expttag{+~\textsc{MINT}}
& 0.893 & 0.956 & 0.727
& 0.833 & 0.957 & 0.641
\\
& \expttag{+~\textsc{STAR}}
& 0.879 & 0.961 & 0.780
& 0.775 & 0.943 & 0.528
\\
\cmidrule(lr){1-8}
\multirow{2}{*}{\textit{\shortstack{Multi\\ple}}}
& \expttag{+~\textsc{SGD}}
& 0.988 & 0.982 & 0.977
& 0.986 & 0.975 & 0.974
\\
& \expttag{+~\textsc{STAR}}
& \textbf{0.998} & \textbf{1.000} & \textbf{0.994}
& \textbf{0.995} & \textbf{0.999} & \textbf{0.983}
\\

\midrule
\multicolumn{2}{l}{\textbf{\textsc{OpenCoder-8B}}}
& 0.713 & 0.870 & 0.546
& 0.715 & 0.909 & 0.505
\\
\cmidrule(lr){1-8}
\multirow{2}{*}{\textit{\shortstack{Single}}}
& \expttag{+~\textsc{MINT}}
& 0.798 & 0.902 & 0.675
& 0.789 & 0.933 & 0.630
\\
& \expttag{+~\textsc{STAR}}
& 0.950 & 0.990 & 0.920
& 0.920 & 0.984 & 0.829
\\
\cmidrule(lr){1-8}
\multirow{2}{*}{\textit{\shortstack{Multi\\ple}}}
& \expttag{+~\textsc{SGD}}
& \textbf{0.997} & \textbf{1.000} & \textbf{0.998}
& 0.996 & \textbf{1.000} & 0.998
\\
& \expttag{+~\textsc{STAR}}
& \textbf{0.997} & 0.998 & \textbf{0.998}
& \textbf{0.997} & \textbf{1.000} & \textbf{0.999}
\\


\midrule
\multicolumn{2}{l}{\textbf{\diff{\textsc{Qwen2.5-Coder-7B}}}}
& 0.716 & 0.866 & 0.554
& 0.729 & 0.928 & 0.520
\\
\cmidrule(lr){1-8}
\multirow{2}{*}{\diff{\textit{\shortstack{Single}}}}
& \expttag{+~\textsc{MINT}}
& 0.863 & 0.926 & 0.774
& 0.856 & 0.958 & 0.715
\\
& \expttag{+~\textsc{STAR}}
& 0.945 & 0.987 & 0.887
& 0.923 & 0.982 & 0.813
\\
\cmidrule(lr){1-8}
\multirow{2}{*}{\diff{\textit{\shortstack{Multi\\ple}}}}
& \expttag{+~\textsc{SGD}}
& \textbf{0.999} & 0.998 & \textbf{0.999}
& \textbf{0.998} & \textbf{1.000} & \textbf{0.998}
\\
& \expttag{+~\textsc{STAR}}
& 0.996 & \textbf{0.999} & 0.978
& 0.993 & \textbf{1.000} & 0.992
\\

\midrule
\multicolumn{2}{l}{\textbf{\diff{\textsc{Qwen2.5-Coder-14B}}}}
& 0.725 & 0.865 & 0.575
& 0.738 & 0.935 & 0.551
\\
\cmidrule(lr){1-8}
\multirow{2}{*}{\diff{\textit{\shortstack{Single}}}}
& \expttag{+~\textsc{MINT}}
& 0.869 & 0.961 & 0.810
& 0.819 & 0.966 & 0.696
\\
& \expttag{+~\textsc{STAR}}
& 0.937 & 0.993 & 0.901
& 0.925 & 0.991 & 0.837
\\
\cmidrule(lr){1-8}
\multirow{2}{*}{\diff{\textit{\shortstack{Multi\\ple}}}}
& \expttag{+~\textsc{SGD}}
& 0.909 & 0.794 & 0.761
& 0.949 & 0.867 & 0.863
\\
& \expttag{+~\textsc{STAR}}
& \textbf{0.993} & \textbf{0.999} & \textbf{0.984}
& \textbf{0.989} & \textbf{0.999} & \textbf{0.989}
\\


\bottomrule

\end{tabular}
 }
\end{table}

On the effectiveness, \textsc{STAR} shows significant improvements over the baselines \textsc{MINT} and \textsc{SGD} in all code generation tasks when repairing almost all LMs.
Compared to \textsc{MINT}, which operates with costly simulation on limited neurons one-by-one, \textsc{STAR} knows the prior information of all neurons to avoid the simulation.
Compared to \textsc{SGD}, which directly utilizes the gradient signals for parameter optimization, \textsc{STAR} uses the prior information to reduce the unnecessary exploration on the loss surface.
Based on the results in \cref{tab:results_rq1_effects}, the effects of repairing methods on LMs appear to vary. On \textsc{CodeLlama-7B} and \textsc{StarCoder2-7B}, \textsc{STAR} shows significant advantages over \textsc{SGD} and \textsc{MINT}. However, the improvements of \textsc{STAR} are very marginal on \textsc{OpenCoder-8B} and \textsc{Qwen2.5-Coder-7B} compared to \textsc{SGD}.
Based on our analysis, \textsc{SGD} tends to converge faster on the LMs whose vocabulary is larger. A larger vocabulary indicates a more powerful tokenization of the data, making the loss surface smoother and easier for \textsc{SGD} to navigate. As shown in \cref{tab:model}, the vocabularies of OpenCoder and Qwen2.5-Coder are $3$ and $5$ times as large as that of \textsc{CodeLlama-7B}. Besides, \textsc{StarCoder2-7B} has a larger vocabulary than \textsc{CodeLlama-7B}, which explains the smaller improvements of \textsc{STAR} over \textsc{SGD} on \textsc{StarCoder2-7B} than on \textsc{CodeLlama-7B}.
In addition, although \textsc{STAR} and \textsc{SGD} have equivalent performance on \textsc{Qwen2.5-Coder-7B}, the former greatly surpasses the latter on \textsc{Qwen2.5-Coder-14B}. This is because the 14B model has a larger dimension of latent space than the 7B model. Our semantic analysis reveals that while higher dimensions make latent representation steering more challenging for \textsc{SGD}, they pose no additional difficulty for \textsc{STAR}. With the prior information introduced in LM optimization, \textsc{STAR} clearly knows the direction to steer representations, but \textsc{SGD} requires time to explore.


\begin{table}[!ht]
    \caption{Efficiency of LM Repair in Coding Tasks}
    \label{tab:results_rq1_times}
    \centering
    \resizebox{0.9\linewidth}{!}{%
        \sisetup{table-format=2.2}
\begin{tabular}{
    ccl rrr r
}

\hiderowcolors
\toprule

\multirow{2}[2]{*}{\textbf{LM}} & \multirow{2}[2]{*}{\textbf{Mode}} & \multirow{2}[2]{*}{\textbf{Method}} & \multicolumn{3}{c}{\textbf{TimeCost (in Seconds) $\downarrow$}} & \multirow{2}[2]{*}{\textbf{Average $\downarrow$}} \\
\cmidrule(lr){4-6}
& & & {\textbf{CoNaLa}} & {\textbf{IA32}} & {\textbf{TLDR}} & \\

\midrule
\showrowcolors

\multirow{4.5}{*}{\rotatebox[origin=c]{90}{\textbf{\shortstack{Code\\Llama-7B}}}}
& \multirow{2}{*}{\rotatebox[origin=c]{90}{\textit{\shortstack{Single}}}}
& \expttag{+~\textsc{MINT}} & 3.117 & 2.629 & 3.656 & 3.134 \\
&& \expttag{+~\textsc{STAR}} & \textbf{1.133} & \textbf{1.025} & \textbf{1.383} & \textbf{1.180} \\
\cmidrule(lr){2-7}
& \multirow{2}{*}{\rotatebox[origin=c]{90}{\textit{\shortstack{Multi\\ple}}}}
& \expttag{+~\textsc{SGD}} & \textbf{0.296} & \textbf{0.362} & \textbf{0.281} & \textbf{0.313} \\
&& \expttag{+~\textsc{STAR}} & 0.391 & 0.547 & 0.404 & 0.447 \\

\midrule
\multirow{4.5}{*}{\rotatebox[origin=c]{90}{\textbf{\shortstack{Star\\Coder2-7B}}}}
& \multirow{2}{*}{\rotatebox[origin=c]{90}{\textit{\shortstack{Single}}}}
& \expttag{+~\textsc{MINT}} & 2.075 & 0.601 & 3.407 & 2.028 \\
&& \expttag{+~\textsc{STAR}} & \textbf{1.079} & \textbf{0.411} & \textbf{1.716} & \textbf{1.069} \\
\cmidrule(lr){2-7}
& \multirow{2}{*}{\rotatebox[origin=c]{90}{\textit{\shortstack{Multi\\ple}}}}
& \expttag{+~\textsc{SGD}} & \textbf{0.342} & \textbf{0.441} & \textbf{0.320} & \textbf{0.368} \\
&& \expttag{+~\textsc{STAR}} & 0.519 & 0.881 & 0.560 & 0.653 \\

\midrule
\multirow{4.5}{*}{\rotatebox[origin=c]{90}{\textbf{\shortstack{Open\\Coder-8B}}}}
& \multirow{2}{*}{\rotatebox[origin=c]{90}{\textit{\shortstack{Single}}}}
& \expttag{+~\textsc{MINT}} & 3.258 & 1.013 & 4.276 & 2.849 \\
&& \expttag{+~\textsc{STAR}} & \textbf{1.150} & \textbf{0.477} & \textbf{1.558} & \textbf{1.062} \\
\cmidrule(lr){2-7}
& \multirow{2}{*}{\rotatebox[origin=c]{90}{\textit{\shortstack{Multi\\ple}}}}
& \expttag{+~\textsc{SGD}} & \textbf{0.336} & \textbf{0.423} & \textbf{0.307} & \textbf{0.355} \\
&& \expttag{+~\textsc{STAR}} & 0.483 & 0.740 & 0.491 & 0.571 \\


\midrule
\multirow{4.5}{*}{\rotatebox[origin=c]{90}{\diff{\textbf{\shortstack{Qwen2.5\\Coder-7B}}}}}
& \multirow{2}{*}{\rotatebox[origin=c]{90}{\diff{\textit{\shortstack{Single}}}}}
& \expttag{+~\textsc{MINT}} & 1.518 & 0.615 & 2.275 & 1.469 \\
&& \expttag{+~\textsc{STAR}} & \textbf{0.724} & \textbf{0.289} & \textbf{1.104} & \textbf{0.706} \\
\cmidrule(lr){2-7}
& \multirow{2}{*}{\rotatebox[origin=c]{90}{\diff{\textit{\shortstack{Multi\\ple}}}}}
& \expttag{+~\textsc{SGD}} & \textbf{0.131} & \textbf{0.184} & \textbf{0.138} & \textbf{0.151} \\
&& \expttag{+~\textsc{STAR}} & 0.266 & 0.487 & 0.323 & 0.359 \\

\midrule
\multirow{4.5}{*}{\rotatebox[origin=c]{90}{\diff{\textbf{\shortstack{Qwen2.5\\Coder-14B}}}}}
& \multirow{2}{*}{\rotatebox[origin=c]{90}{\diff{\textit{\shortstack{Single}}}}}
& \expttag{+~\textsc{MINT}} & 1.590 & 0.601 & 2.183 & 1.458 \\
&& \expttag{+~\textsc{STAR}} & \textbf{1.223} & \textbf{0.480} & \textbf{1.833} & \textbf{1.179} \\
\cmidrule(lr){2-7}
& \multirow{2}{*}{\rotatebox[origin=c]{90}{\diff{\textit{\shortstack{Multi\\ple}}}}}
& \expttag{+~\textsc{SGD}} & \textbf{0.193} & \textbf{0.283} & \textbf{0.218} & \textbf{0.231} \\
&& \expttag{+~\textsc{STAR}} & 0.438 & 0.824 & 0.545 & 0.602 \\



\bottomrule

\end{tabular}

 }
\end{table}

On the efficiency, as shown in \cref{tab:results_rq1_times}, \textsc{STAR} realizes a 2.4--7.0 times speedup on average compared to \textsc{MINT}, but is slower than \textsc{SGD} slightly.
The main reason is that: \textsc{MINT} requires multiple forward computation for sufficient simulation; \textsc{SGD} is the minimal solution of using gradient signals, so any methods relying on the gradients cannot be faster, so does \textsc{STAR}.
%
The gap in efficiency between \textsc{MINT} and \textsc{STAR} is significant, which is mainly caused by the differences in the required amount of LM forward or backward computations.
\textsc{MINT} is a pipeline consisting of individual steps for neuron attribution, patch estimation, and failure localization. In contrast, \textsc{STAR} only requires two steps: one is neuron attribution, while the other one is patch estimation. The failure localization of \textsc{MINT} takes a long time since it does multiple forward computations in addition to measure and compare the gains of patching each neuron.
Meanwhile, \textsc{STAR} merges the required forward computations in attribution and estimation to reduce time cost.
Since the main time cost of \textsc{SGD} is for gradients computation, \textsc{SGD} is the minimal solution of parameter optimization, and faster than any other methods using gradients. For example, \textsc{STAR} relies on gradients, so it takes a same time cost for computing gradients, while \textsc{STAR} has an additional computation for solving the ``prior'' information. It explains why \textsc{STAR} must be slower than \textsc{SGD}.


Comparing the two modes for LM repair, in most models and datasets, \textit{multiple}-mode methods perform better than \textit{single}-mode, since \textit{multiple}-mode incorporates multiple repairs into one, reducing their mutual interferences. However, it is not true for \textsc{STAR} in repairing \textsc{CodeLlama-7B} on the IA32 data. This is because the two modes correspond to two modes of next-token prediction.
The \textit{multiple}-mode computes ``prior'' using the state of model parameters before repairing multiple failures, while the \textit{single}-mode computes ``prior'' using the state before repairing only one failure. When repairing continuous failures, the prior info of \textit{multiple}-mode is more likely to become inaccurate than that of \textit{single}-mode.
In addition, \textit{single}-mode methods take more time than \textit{multiple}-mode methods. The former processes each failure separately, while the latter processes multiple failures together. By combining repeated operations in LM repair, the time cost is notably reduced.

\begin{table}[!ht]
    \caption{Overfitting Risk Assessment of LM Repair (CoNaLa)}
    \label{tab:results_rq1_of}
    \centering
    \resizebox{0.9\linewidth}{!}{%
        \sisetup{table-format=2.2}
\begin{tabular}{
    ccl rrrr
}

\hiderowcolors
\toprule

\multirow{2}[2]{*}{\textbf{LM}} & \multirow{2}[2]{*}{\textbf{Mode}} & \multirow{2}[2]{*}{\textbf{Method}} & \multicolumn{3}{c}{\textbf{BLEU-RPD $\downarrow$}} & \multirow{2}[2]{*}{\textbf{\shortstack{Degradation\\Percentage $\downarrow$}}} \\
\cmidrule(lr){4-6}
& & & {\textbf{Mean}} & {\textbf{Std}} & {\textbf{IQR}} & \\

\midrule
\multirow{4.5}{*}{\rotatebox[origin=c]{90}{\textbf{\shortstack{Qwen2.5\\Coder-7B}}}}
& \multirow{2}{*}{\rotatebox[origin=c]{90}{\textit{\shortstack{Single}}}}
& \expttag{+~\textsc{MINT}} & \textbf{0.000} & \textbf{0.004} & \textbf{0.003} & \textbf{27.75\%} \\
&& \expttag{+~\textsc{STAR}} & 0.001 & \textbf{0.004} & 0.006 & 54.25\% \\
\cmidrule(lr){2-7}
& \multirow{2}{*}{\rotatebox[origin=c]{90}{\textit{\shortstack{Multi\\ple}}}}
& \expttag{+~\textsc{SGD}} & 0.011 & 0.026 & 0.016 & 70.00\% \\
&& \expttag{+~\textsc{STAR}} & \textbf{0.000} & \textbf{0.005} & \textbf{0.006} & \textbf{43.00\%} \\

\midrule
\multirow{4.5}{*}{\rotatebox[origin=c]{90}{\textbf{\shortstack{Qwen2.5\\Coder-14B}}}}
& \multirow{2}{*}{\rotatebox[origin=c]{90}{\textit{\shortstack{Single}}}}
& \expttag{+~\textsc{MINT}} & 0.030 & 0.097 & 0.014 & \textbf{62.25\%} \\
&& \expttag{+~\textsc{STAR}} & \textbf{0.003} & \textbf{0.005} & \textbf{0.005} & 66.75\% \\
\cmidrule(lr){2-7}
& \multirow{2}{*}{\rotatebox[origin=c]{90}{\textit{\shortstack{Multi\\ple}}}}
& \expttag{+~\textsc{SGD}} & 0.197 & 0.320 & 0.120 & 97.50\% \\
&& \expttag{+~\textsc{STAR}} & \textbf{0.002} & \textbf{0.006} & \textbf{0.007} & \textbf{56.50\%} \\

\bottomrule

\end{tabular}

 }
\end{table}

\paragraph{Overfitting Risk Assessment}
Based on the results in \cref{tab:results_rq1_of}, \textsc{SGD} already causes overfitting while LM repair methods show no overfitting risk.
Compared with others, \textsc{SGD} has notably larger statistical values and degradation percentage. It means that SGD-based parameter optimization causes LM to underperform more often, and with more severe degradation. The reason why \textsc{SGD} performs badly is due to the difficulty in fitting individual failures without damaging the normal LM firing pattern.
In contrast, repairing methods' statistical values are almost zero, and their degradation percentage is very healthy, smaller than or close enough to $50\%$. The reason is that, \textsc{MINT} and \textsc{STAR} introduce ``prior'' to update parameters, and they target buggy neurons leveraging attribution. \textsc{MINT} patches one neuron each time while \textsc{STAR} relies on sparsity patterns, so \textsc{MINT} has even better results than \textsc{STAR}.
Besides, for \textsc{STAR}, the \textit{multiple}-mode is better, since solving a complex patch for multiple failures tends to eliminate the compounding errors of repeated repairs.
Comparing 7B and 14B models, larger LMs seem more likely to undergo overfitting, while \textsc{STAR} remains a low overfitting risk.


\subsection{\req{2} Results}
\label{subsec:rq2_results}

When converting LM repair to the form of parameter optimization, targeting the critical neurons guarantees better performance than optimizing all parameters.
As a whole, the results of the sparsity pattern indicate the importance of ``targeting'' in LM repair.

\begin{figure}[!ht]
    \centering
    \includegraphics[width=\linewidth]{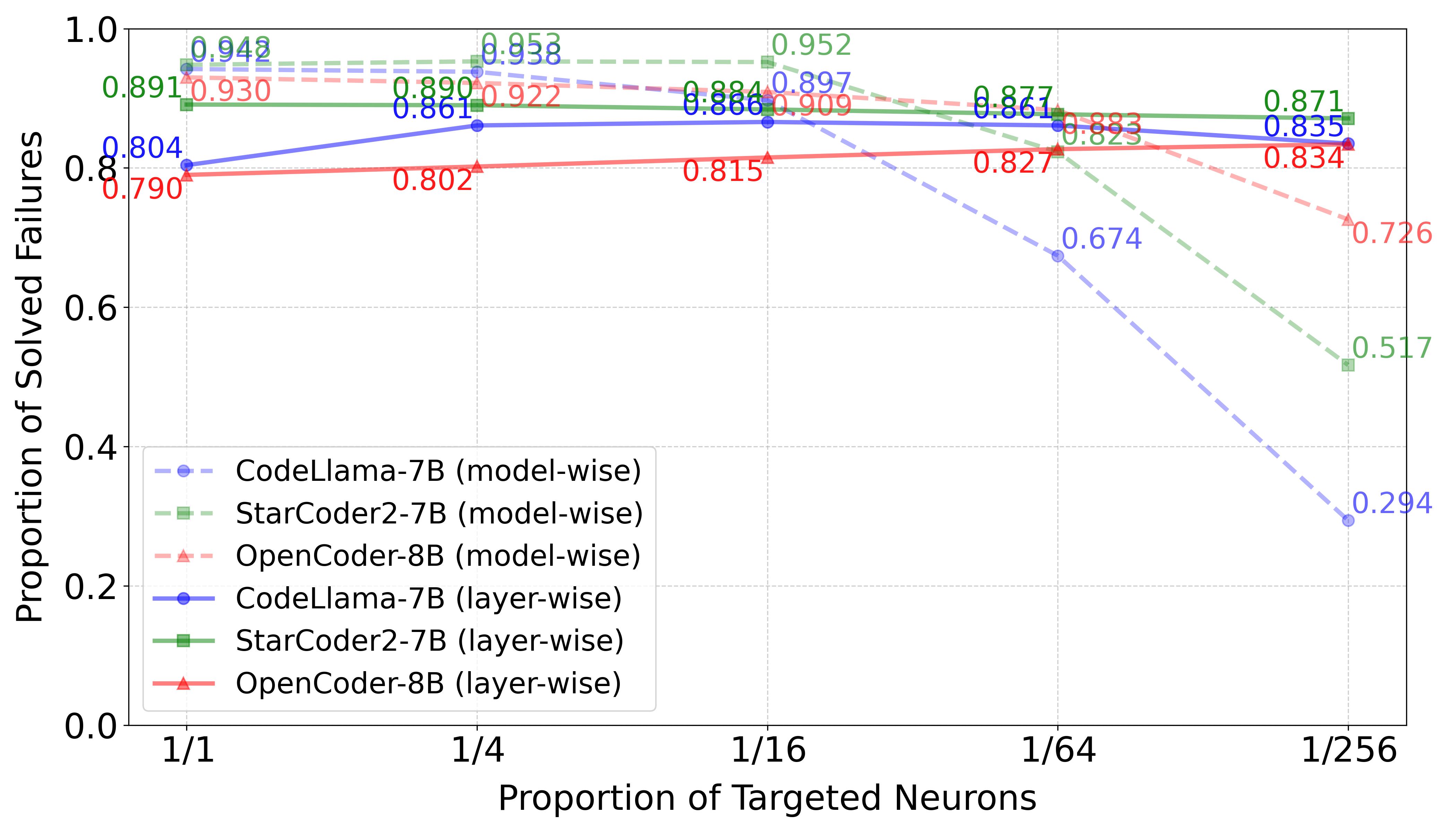}
    \caption{Accuracy of LMs and repaired LMs on CoNaLa, with model-wise and layer-wise sparsity pattern.}
    \label{fig:pattern_conala}
\end{figure}

% ...
First, the multi-layered hierarchy prefers the layer-wise sparsity pattern, instead of the model-wise sparsity pattern.
As shown in \cref{fig:pattern_conala}, when the proportion of targeted neurons becomes smaller, the proportion of solved failures remains at a high level for layer-wise \textsc{STAR}.
In contrast, the performance will face huge drops for model-wise \textsc{STAR}, especially when the proportion of targeted neurons becomes smaller than $\frac{1}{16}$.
All three models undergo similar patterns of change. When the proportion of targeted neurons is close to $100\%$, layer-wise \textsc{STAR} is not as good as model-wise \textsc{STAR} initially but will gradually outperform the latter when the proportion becomes smaller and smaller.

\begin{figure}[!ht]
    \centering
    \includegraphics[width=\linewidth]{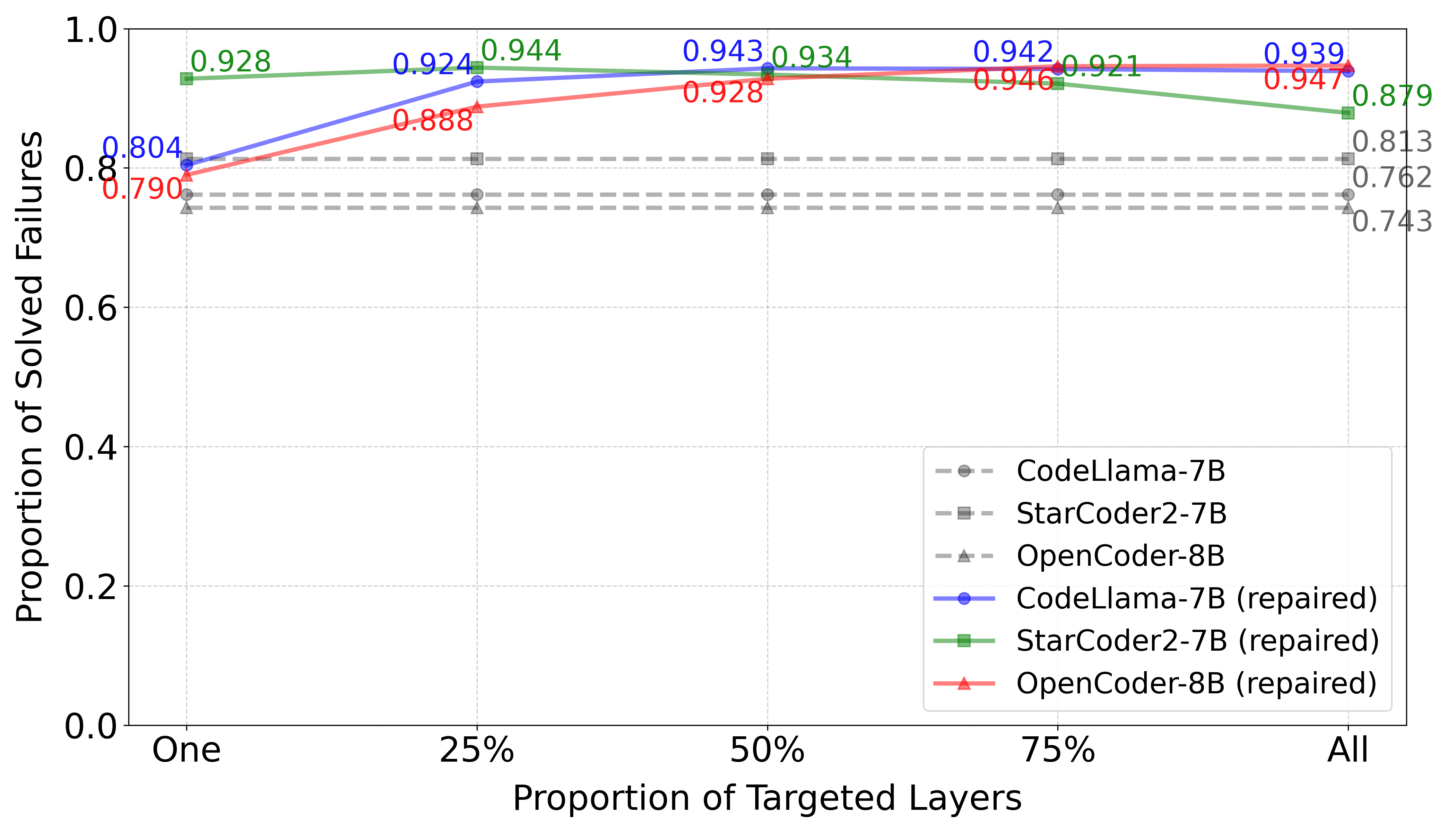}
    \caption{Accuracy of LMs and repaired LMs on CoNaLa, with a varying proportion of targeted layers.}
    \label{fig:layer_conala}
\end{figure}

% ...
Then, the proportion of targeted layers affects the performance. Prioritizing the layers that cause the most change to the loss, when the proportion is close to half, or slightly higher, the performance can be better.
As shown in \cref{fig:layer_conala}, comparing the performance of each model with that of its corresponding repaired model, we see that no matter the proportion of targeted layers, LM repair always guarantees an improved proportion of solved failures.
For different models, the patterns of change show slight differences. \textsc{CodeLlama-7B} and \textsc{OpenCoder-8B} undergo a slowly rising trend, while \textsc{StarCoder2-7B} undergoes a slowly declining trend.

\begin{figure}[!ht]
    \centering
    \includegraphics[width=\linewidth]{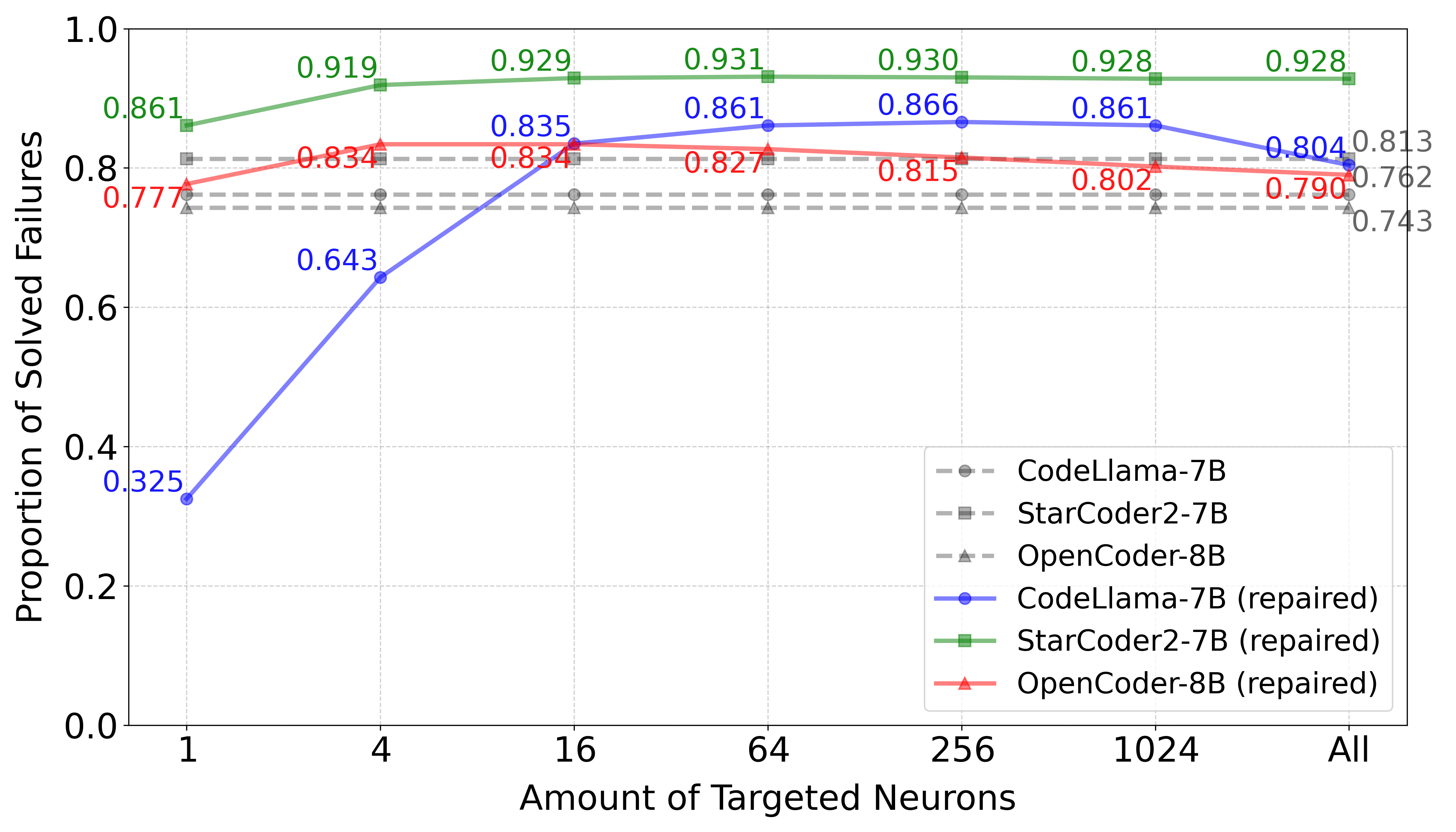}
    \caption{Accuracy of LMs and repaired LMs on CoNaLa, with a varying amount of targeted neurons.}
    \label{fig:neuron_conala}
\end{figure}

% ...
Third, the amount of target neurons affects the performance. Specifying the layer that is most critical to the loss, when the optimal amount is around $64$, or $\frac{1}{16}$ of the number of neurons in each layer, the performance is optimal.
As shown in \cref{fig:neuron_conala}, when the amount of targeted neurons is too small, such as \textsc{CodeLlama-7B}, the model even performs better than the corresponding repaired one. Meanwhile, the repaired model will face a slight decrease in its performance.
In contrast, the performance of \textsc{StarCoder2-7B} and \textsc{OpenCoder-8B} remains stable, remaining higher than the model performance when without being repaired.

\subsection{\req{3} Results}
\label{subsec:rq3_results}


In \req{3}, we evaluate the side effects of LM methods by measuring both the generalization and specificity of the repaired LMs, with MAE/RMSE scores and their harmonic-mean value (GSH scores). High scores indicate good performance in the respective aspects, and a high GSH score indicates a good balance between them.

\begin{table}[!htb]
    \caption{Side Effects of LM Repair in Coding Tasks}
    \label{tab:results_rq3_reliability}
    \centering
    \resizebox{\linewidth}{!}{%
        \sisetup{table-format=2.2}
\begin{tabular}{
    cl rrr rrr
}

\toprule

\multirow{2}[2]{*}{\textbf{LM}} & \multirow{2}[2]{*}{\textbf{Method}} & \multicolumn{2}{c}{\textbf{Generalization $\uparrow$}} & \multicolumn{2}{c}{\textbf{Specificity $\uparrow$}} & \multicolumn{2}{c}{\textbf{GSH $\uparrow$}} \\
\cmidrule(lr){3-4} \cmidrule(lr){5-6} \cmidrule(lr){7-8}
&
& \textbf{$\text{MAE}$} & \textbf{$\text{RMSE}$}
& \textbf{$\text{MAE}$} & \textbf{$\text{RMSE}$}
& \textbf{$\text{MAE}$} & \textbf{$\text{RMSE}$} \\


\midrule

\multirow{4.5}{*}{\rotatebox[origin=c]{90}{\textbf{\shortstack{Code\\Llama-7B}}}}
& \expttag{+~\textsc{MINT}}
& 0.007 & 0.081
& \tabh 0.993 & \tabh 0.919
& $0.014$ & 0.149 \\
& \expttag{+~\textsc{SGD}}
& \tabh 0.988 & \tabh 0.990
& 0.039 & 0.030
& 0.075 & 0.058 \\
& \expttag{+~\textsc{STAR}}
& 0.925 & 0.954
& 0.114 & 0.069
& 0.203 & 0.129 \\
& \exptlabel{+~\textsc{STAR}}{layer}
& 0.459 & 0.614
& \tabh 0.586 & \tabh 0.426
& \textbf{0.515} & \textbf{0.503} \\
& \exptlabel{+~\textsc{STAR}}{neuron}
& 0.011 & 0.063
& \tabh 0.987 & \tabh 0.944
& 0.022 & 0.118 \\

\midrule

\multirow{4.5}{*}{\rotatebox[origin=c]{90}{\textbf{\shortstack{Star\\Coder2-7B}}}}
& \expttag{+~\textsc{MINT}}
& 0.000 & 0.000
& \tabh 1.000 & \tabh 1.000
& $0.000$ & $0.000$ \\
& \expttag{+~\textsc{SGD}}
& \tabh 0.959 & \tabh 0.971
& 0.336 & 0.261
& 0.498 & 0.411 \\
& \expttag{+~\textsc{STAR}}
& 0.929 & 0.952
& 0.683 & 0.510
& \textbf{0.787} & 0.664 \\
& \exptlabel{+~\textsc{STAR}}{layer}
& 0.836 & 0.897
& \tabh 0.739 & \tabh 0.563
& 0.785 & \textbf{0.692} \\
& \exptlabel{+~\textsc{STAR}}{neuron}
& 0.017 & 0.052
& \tabh 0.978 & \tabh 0.943
& 0.033 & 0.099 \\

\midrule

\multirow{4.5}{*}{\rotatebox[origin=c]{90}{\textbf{\shortstack{Open\\Coder-8B}}}}
& \expttag{+~\textsc{MINT}}
& 0.001 & 0.027
& \tabh 1.000 & \tabh 0.999
& 0.002 & 0.053 \\
& \expttag{+~\textsc{SGD}}
& \tabh 0.904 & \tabh 0.928
& 0.525 & 0.429
& \textbf{0.664} & 0.587 \\
& \expttag{+~\textsc{STAR}}
& 0.378 & 0.555
& 0.891 & 0.754
& 0.531 & \textbf{0.639} \\
& \exptlabel{+~\textsc{STAR}}{layer}
& 0.054 & 0.126
& \tabh 0.971 & \tabh 0.913
& 0.102 & 0.221 \\
& \exptlabel{+~\textsc{STAR}}{neuron}
& 0.006 & 0.014
& \tabh 0.996 & \tabh 0.986
& 0.012 & 0.028 \\

\bottomrule

\end{tabular}

}
\end{table}

Based on the GSH scores in \cref{tab:results_rq3_reliability}, our approach \textsc{STAR} is performant in most comparisons. In \textsc{StarCoder2-7B} and \textsc{OpenCoder-8B}, its GSH scores are significantly higher than others (and separately comparable to one variant and \textsc{SGD}). Even though in \textsc{CodeLlama-7B}, its GSH scores are second only to its variant \exptlabel{\textsc{STAR}}{layer}. This proves the advantages of our approach in reducing the side effects of LM repair, or realizing a better balance between generalization and specificity. The changes in its GSH scores in different models show the points of future improvements that, the effects of \textsc{STAR} cannot well guarantee stability yet.
The baseline \textsc{SGD} is not leading concerning the balance between generalization and specificity.
However, if comparing the magnitudes of its generalization scores, we see \textsc{SGD} likely to have larger values than other methods, especially in \textsc{OpenCoder-8B}. It shows the unique advantages of its great generalization.
Meanwhile, its GSH scores are rather low in \textsc{CodeLlama-7B}, not consistent with its performance in the other two models. The phenomenon confirms our conclusion again that, the effects of optimization methods cannot guarantee the stability in different models.
\textsc{MINT} performs like an extreme case, since $4$ out of its $6$ GSH scores are lowest in comparisons. \textsc{MINT} has very low generalization scores but the highest specificity scores. It shows that \textsc{MINT} has limitations in guaranteeing its generalization even though its specificity is already outstanding.

Comparing two \textsc{STAR} variants, \exptlabel{\textsc{STAR}}{layer} shows obviously better balance than \exptlabel{\textsc{STAR}}{neuron} in all three models. Based on our analysis, this is caused by the amount of neurons being patched. The former takes around hundreds times more than the latter. It suggests that, a good sparsity pattern of LM repair in improving LM performance cannot promise LMs good side effects. Moreover, when reducing the number of neurons in LM repair, compared to specificity which is already at a very good level, generalization is more likely to be affected to a very bad level. It further indicates that, the main challenge of LM repair methods is improving the generalization instead of improving the specificity.


\paragraph{\textbf{Generalization Evaluation}}
As shown in \cref{tab:results_rq3_reliability}, compared to \textsc{STAR}, in all three models, \textsc{SC} is the only method showing stably higher generalization scores. That is, \textsc{STAR} and \textsc{SC} show consistently larger positive effects (to the related data) than other methods.
Besides, \textsc{MINT} and \exptlabel{\textsc{STAR}}{neuron} have similar performance and perform the worst in all models.
\exptlabel{\textsc{STAR}}{layer} shows inconsistent performance, with results varying significantly across different models. It shows competitive performance to \textsc{STAR} in \textsc{StarCoder2-7B}, but shows huge gaps to \textsc{STAR} in the other two models.
Based on our analysis, the sparsity pattern led to the performance drop, and the attribution method is responsible for the inconsistency.

\paragraph{\textbf{Specificity Evaluation}}
As shown in \cref{tab:results_rq3_reliability}, compared to \textsc{STAR}, in all three models, \textsc{MINT} and two variants have stably higher specificity scores, indicating their better specificity. In contrast, \textsc{SGD} has the lowest specificity scores, indicating its consistently larger negative effects (on the unrelated data).
On one hand, LM repair methods are likely to have overall better specificity than common optimization methods, such as \textsc{SGD}; on the other hand, as mentioned, the use of sparsity patterns does show effects in improving the specificity of \textsc{STAR} in LM repair.
Further, we see the strong correlations between generalization and specificity, since it is challenging to attain high scores in both aspects concurrently.

\begin{table}[!ht]
    \caption{Cumulative Impact Assessment of LM Repair}
    \label{tab:results_discuss_cc}
    \centering
    \resizebox{\linewidth}{!}{%
        \sisetup{table-format=2.2}
\begin{tabular}{
    ll rr rr
}

\toprule

\multirow{2}[2]{*}{\textbf{Mode}} & \multirow{2}[2]{*}{\textbf{Method}} & \multicolumn{2}{c}{\textbf{ExactMatch $\uparrow$}} & \multicolumn{2}{c}{\textbf{BLEU $\uparrow$}} \\
\cmidrule(lr){3-4} \cmidrule(lr){5-6}
&
& {\textbf{HumanEval}} & {\textbf{BigCode}}
& {\textbf{HumanEval}} & {\textbf{BigCode}}
\\

\midrule
\multicolumn{2}{l}{\textbf{\textsc{Qwen2.5-Coder-7B}}}
& 0.816 & 0.804
& 0.791 & 0.734
\\
\cmidrule(lr){1-6}
\multirow{2}{*}{\textit{\shortstack{Single}}}
& \expttag{+~\textsc{MINT}}
& 0.881 & 0.870
& 0.861 & 0.819
\\
& \expttag{+~\textsc{STAR}}
& 0.942 & 0.909
& 0.911 & 0.855
\\
\cmidrule(lr){1-6}
\multirow{2}{*}{\textit{\shortstack{Multi\\ple}}}
& \expttag{+~\textsc{SGD}}
& 0.987 & 0.923
& \textbf{0.966} & 0.886
\\
& \expttag{+~\textsc{STAR}}
& \textbf{0.993} & \textbf{0.981}
& 0.958 & \textbf{0.959}
\\

\midrule
\multicolumn{2}{l}{\textbf{\textsc{Qwen2.5-Coder-14B}}}
& 0.823 & 0.807
& 0.799 & 0.738
\\
\cmidrule(lr){1-6}
\multirow{2}{*}{\textit{\shortstack{Single}}}
& \expttag{+~\textsc{MINT}}
& 0.844 & 0.817
& 0.782 & 0.749
\\
& \expttag{+~\textsc{STAR}}
& 0.942 & 0.885
& 0.931 & 0.826
\\
\cmidrule(lr){1-6}
\multirow{2}{*}{\textit{\shortstack{Multi\\ple}}}
& \expttag{+~\textsc{SGD}}
& \textbf{0.981} & \textbf{0.965}
& \textbf{0.970} & \textbf{0.951}
\\
& \expttag{+~\textsc{STAR}}
& 0.971 & 0.896
& 0.917 & 0.844
\\


\bottomrule

\end{tabular}
}
\end{table}

\paragraph{Cumulative Impact Assessment}
Based on the results in \cref{tab:results_discuss_cc}, LM repair methods smoothly scale to LMs of varying sizes. \textsc{STAR} shows significant improvements in the LM, even though it appears less performant than \textsc{SGD}. Besides \textsc{STAR}, \textsc{MINT} also shows a reliable capability of LM repair.
Comparing the results of the 7B model with 14B, \textsc{SGD} performs stably on HumanEval while showing further improvements on BigCode. This is different from the finding of performance degradation in RQ1. The distinction correlates with the amount of failures to solve. On average, each HumanEval data contains fewer failures than BigCode but more than the datasets in RQ1 (refer to \cref{subsec:dataset}).
On the cumulative impact (by comparing the results of HumanEval and BigCode), repairing methods guarantee the model's overall stability after solving around $50$ failures since the repaired LMs still perform better. However, the repairing effects show diminishing returns, especially for the 14B model.
Although \textsc{MINT} shows marginal diminishing returns, \textsc{STAR} has better performance on BigCode, especially the \textit{multiple}-mode.
In addition, \textsc{SGD} shows similar diminishing returns, which suggests the correlations with the increased difficulty of LM optimization.

\section{Discussion}
\label{sec:discussion}



To demonstrate the effects of LM repair, we select an actual and representative case from the experiments, as shown in \cref{tab:showcase}.
It is the $32\-th$ sample in TDLR, the natural language prompt is ``\textit{submit a job and request multiple nodes}'', and the correcting ground truth is ``\textit{sbatch --nodes=\{\{3\}\} \{\{path/to/job.sh\}\}}''.
The incorrect tokens in the sequences are highlighted in red while others are in blue.

Compared with the originally generated results which show $4$ failures, the results of repaired LMs are better. The \textit{multiple}-mode methods solve more failures than \textit{single}-mode methods.
However, we see repairing methods suffer from a common issue that, LM repair may not generalize well as intended. In particular, it is hard to avoid the current repair affecting normal model behaviors.

Even though all methods successfully solved the model failure that wrongly predicted ``job'' as ``script'', \textsc{MINT} and \textsc{STAR} caused a new failure that, a token earlier in the sequence was wrongly predicted as ``job''.
In this paper, we proposed an optimization method for LM repair, which is compatible with the \textit{multiple}-mode to mitigate this issue. However, when processing multiple failures together, it is tricky to adjust their weights dynamically in a flexible manner.




\begin{table}[!ht]
    \caption{Showcase of Multiple Repairs in Succession.}
    \label{tab:showcase}
    \centering
    \resizebox{\linewidth}{!}{%
        \sisetup{table-format=2.2}
\rowcolors{2}{}{gray!10}
\begin{tabular}{
    cll
}

\hiderowcolors
\toprule

\textbf{Mode} & \textbf{Method} & \textbf{String and Token Sequence} \\

\midrule



\multicolumn{2}{c}{\textbf{\textsc{StarCoder2-7B}}} & \textit{\hlpink{q}~\hlcyan{batch}~\hlcyan{ --}~\hlcyan{nodes}~\hlcyan{=\{\{}~\hlpink{2}~\hlcyan{\}\}}~\hlcyan{ \{\{}~\hlpink{script}~\hlcyan{/}~\hlcyan{to}~\hlcyan{/}~\hlpink{script}~\hlcyan{.}~\hlcyan{sh}~\hlcyan{\}\}}} \\
\cmidrule(lr){1-3}
\multirow{2}{*}{\rotatebox[origin=c]{90}{\textit{\shortstack{Single}}}}
& \expttag{+~\textsc{MINT}} & \textit{\hlpink{q}~\hlcyan{batch}~\hlcyan{ --}~\hlcyan{nodes}~\hlcyan{=\{\{}~\hlpink{2}~\hlcyan{\}\}}~\hlcyan{ \{\{}~\hlpink{job}~\hlcyan{/}~\hlcyan{to}~\hlcyan{/}~\hlcyan{job}~\hlcyan{.}~\hlcyan{sh}~\hlcyan{\}\}}} \\
& \expttag{+~\textsc{STAR}} & \textit{\hlcyan{s}~\hlcyan{batch}~\hlcyan{ --}~\hlpink{job}~\hlpink{=}~\hlcyan{3}~\hlcyan{\}\}}~\hlpink{ --}~\hlcyan{path}~\hlcyan{/}~\hlcyan{to}~\hlcyan{/}~\hlcyan{job}~\hlcyan{.}~\hlcyan{sh}~\hlcyan{\}\}}} \\
\cmidrule(lr){1-3}
\multirow{2}{*}{\rotatebox[origin=c]{90}{\textit{\shortstack{Multi\\ple}}}}
& \expttag{+~\textsc{SGD}} & \textit{\hlcyan{s}~\hlcyan{batch}~\hlcyan{ --}~\hlcyan{nodes}~\hlcyan{=\{\{}~\hlcyan{3}~\hlcyan{\}\}}~\hlcyan{ \{\{}~\hlcyan{path}~\hlcyan{/}~\hlcyan{to}~\hlcyan{/}~\hlcyan{job}~\hlcyan{.}~\hlcyan{sh}~\hlcyan{\}\}}} \\
& \expttag{+~\textsc{STAR}} & \textit{\hlcyan{s}~\hlcyan{batch}~\hlcyan{ --}~\hlcyan{nodes}~\hlcyan{=\{\{}~\hlcyan{3}~\hlcyan{\}\}}~\hlcyan{ \{\{}~\hlpink{job}~\hlcyan{/}~\hlcyan{to}~\hlcyan{/}~\hlcyan{job}~\hlcyan{.}~\hlcyan{sh}~\hlcyan{\}\}}} \\

\bottomrule

\end{tabular}

 }
\end{table}

\section{Related Work}
\label{sec:related_work}

Neural network (NN) repair is a well-studied area in software engineering, but existing work mainly targets CNNs and discriminative tasks, making it challenging to adapt to language models and generative tasks~\cite{Sohn2019ArachneSR,Gao2022FairneuronID}.
The common strategies for repairing neural networks involve identifying faulty neurons via causal analysis~\cite{Sun2022CausalityBasedNN} and updating them with adaptive methods~\cite{LiCalsi2023AdaptiveSR}.
Also, using an additional inﬂuence model to characterize the stateful and statistical behaviors of the target model to interpret and repair the incorrect behaviors of RNNs~\cite{Xie2021RNNRepairAR}.
The recent approaches, such as \textsc{VeRe}~\cite{Ma2024VereVG}, have shown promise in repairing CNNs, particularly for backdoor removal, even though verification-based techniques may have to face high computational costs.
The topic of automated repair for LMs has not yet been thoroughly explored. The prior work on NN repair is not directly applicable to language models.
%
To the best of our knowledge, the state-of-the-art \textsc{MINT} is the pioneering work of LM repair. It formalizes the problem of LM repair as solving LM failures, and provides an intuitive solution leveraging the semantic property of LMs, which is effective, and efficient, while showing tolerable side effects.
\textsc{MINT} is a pipeline that recognize and solve LM failure at the neuron-level.
Inspired by \textsc{MINT}, \textsc{STAR} proposes a complete semantic-based framework and realizes an optimization method that further improves the performance of LM repair and reduces the side effects.
Unlike \textsc{MINT}, which follows a sequential pipeline of three individual stages, \textsc{STAR} formulates LM repair as an unified optimization. Therefore, compared to \textsc{MINT}, \textsc{STAR} supports important features, including: proving its efficacy theoretically (leveraging the loss surface); solving multiple LM failures in a single run; and applying a sparsity pattern to filter buggy neurons. In addition, even though both \textsc{STAR} and \textsc{MINT} rely on semantics-related concepts~\cite{gu2024vds,gu2025salf}, their underlying beliefs are different: \textsc{STAR} advocates that neurons and knowledge follow many-to-many dynamic associations~\cite{allen-zhu2025physics}, in contrast to MINT's controversial one-to-many associations~\cite{wei2024does}.



Model editing technique indicates a wide range of methods~\cite{Yao2023EditingLL}.
Compared to model editing methods, especially those directly modifying model parameters without using additional modules or components, LM repair seems similar in methodology. However, LM repair focuses on targeting certain neurons with tailored updates while remaining computationally lightweight.
In general, model editing methods require a large corpus as additional data to build an initial state of model parameters, trying to cover the most common subset of the LM's pretrained knowledge. They are data-driven, while LM repair directly leverages the inherent property of LMs, such as the semantic property in the LM latent space.
Besides, they aim to modify factual knowledge stored in LMs for certain knowledge tasks, while LM repair aims to fix incorrect behaviors, which may not be directly related to factual knowledge~\cite{Zhang2024ACS}. It means the parameters to optimize in LM repair are not constrained in certain modules.
For the need of repairing LMs, model editing methods lack enough feasibility and interpretability, which limit their usefulness~\cite{He2025KnowledgeUN}.
The typical methods like ROME~\cite{Meng2022LocatingAE} and MEMIT~\cite{Meng2022MassEditingMI} identify critical model layers and then use gradients to update parameters. ROME locates key neuron activations through causal interventions, updating specific facts by modifying FFN weights in intermediate layers. MEMIT extends this by enabling batch edits, allowing simultaneous updates to multiple facts across model layers.
The recent methods further refine these approaches. PMET~\cite{Li2023PMETPM} considers interactions between MHA, FFN, and residual connections, optimizing both MHA and FFN for better knowledge updates. AlphaEdit~\cite{Fang2024AlphaEditNC} minimizes disruption by projecting changes onto the null space of preserved knowledge, ensuring original information remains intact. These innovations address earlier limitations and enhance the precision of model updates.


\section{Conclusion}
\label{sec:conclusion}

In this paper, we propose a semantic-based optimization framework for automated repair of LMs.
LM repair is the process of solving LM failures by targeting and tailoring updates.
By converting a ``locate-and-patch'' pipeline to an optimization process, common limitations and doubts of LM repair can be mitigated.
Leveraging the semantics of LMs, our framework \textsc{STAR} suggests solving the updates to model parameters based on the intended steering of latent representations, which are the required changes in geometric position from the current logits to the ground truth.
Further, the deltas serve as prior guidance in optimization, and the sparsity pattern of LMs enables the updates targeting certain neurons.
% ...
The results of our extensive experiments show that \textsc{STAR} outperforms the prior work in both performance and side effects.

In our future work, we will study more on the sparsity pattern of LMs, as well as the effects of layer-wise hierarchy in LM repair. By doing this, LM repair techniques will become more flexible and applicable to real-world tasks.
Meanwhile, we plan to explore the uses of LM repair as an intervention tool for the underlying thought processes of LMs, focusing on challenging coding tasks~\cite{Liu2023CodeEW,Chen2024ReasoningRB}.





\begin{acks}
This project was funded by the \grantsponsor{arc}{Australian Research Council (ARC)}{https://www.arc.gov.au/}, Discovery Project \grantnum{arc}{DP210100041}. We thank \grantsponsor{arc}{National Computational Infrastructure (NCI)}{https://nci.org.au/} for providing the computing resources.
\end{acks}

\bibliographystyle{ACM-Reference-Format}
\bibliography{references}



\end{document}